\documentclass{aa}  

\usepackage[export]{adjustbox}
\usepackage{subfig}
\usepackage[T1]{fontenc}
\usepackage{ae,aecompl}
\usepackage{amsmath}	
\usepackage{amssymb}	
\usepackage{color}
\usepackage{graphicx}
\usepackage{txfonts}
\usepackage{hyperref}
\usepackage{tikz}
\usepackage{colortbl}
\newcolumntype{P}[1]{>{\centering\arraybackslash}p{#1}}

\begin{document} 
\newcommand{\dd}{deg$^{2}$}

\title{The XXL survey}

\subtitle{ XLVI: Forward cosmological analysis of the C1 cluster sample\thanks{Based on observations obtained with XMM-Newton, an ESA science mission with instruments and contributions directly funded by ESA Member States and NASA}}

\author{Christian Garrel
          \inst{1}, 
          Marguerite Pierre \inst{1},
          Patrick Valageas \inst{2},
          Dominique Eckert \inst{3},
          Federico Marulli \inst{4,5,6},
          Alfonso Veropalumbo \inst{7},
          Florian Pacaud \inst{8}, 
          Nicolas Clerc \inst{9},
          Mauro Sereno \inst{5,6},
          Keiichi Umetsu \inst{10},
          Lauro Moscardini \inst{4,5,6},
          Sunayana Bhargava \inst{1}, 
          Christophe Adami \inst{11},
          Lucio Chiappetti \inst{12},
          Fabio Gastaldello \inst{12},
          Elias Koulouridis \inst{13},
          Jean-Paul Le Fevre \inst{14},
          Manolis Plionis \inst{15,16}
          }
\institute{
   AIM, CEA, CNRS, Universit\'e Paris-Saclay, Universit\'e Paris Diderot, Sorbonne Paris Cite, F-91191 Gif-sur-Yvette, France\\
              \email{christian.garrel@cea.fr}
         \and
    Institut de Physique Théorique, CEA Saclay, F-91191 Gif-sur-Yvette, France
        \and
    Department of Astronomy, University of Geneva, ch. d'Écogia 16, CH-1290 Versoix Switzerland
        \and
    Dipartimento di Fisica e Astronomia "Augusto Righi" - Alma Mater Studiorum Università di Bologna, via Piero Gobetti 93/2, I-40129 Bologna, Italy
        \and
    INAF - Osservatorio di Astrofisica e Scienza dello Spazio di Bologna, via Piero Gobetti 93/3, I-40129 Bologna, Italy
        \and
    INFN, Sezione di Bologna, viale Berti Pichat 6/2, I-40127 Bologna, Italy
        \and
    Dipartimento di Fisica, Università degli Studi Roma Tre, via della Vasca Navale 84, I-00146 Rome, Italy
        \and
    Argelander Institut für Astronomie, Universität Bonn, D-53121 Bonn, Germany
        \and
    IRAP, Université de Toulouse, CNRS, UPS, CNES, F-31028 Toulouse, France
        \and
    Academia Sinica Institute of Astronomy and Astrophysics (ASIAA), No. 1, Section 4, Roosevelt Road, Taipei 10617, Taiwan
        \and
    Aix Marseille Univ., CNRS, CNES, LAM, Marseille, France
        \and
    INAF - IASF Milan, via A. Corti 12, I-20133 Milano, Italy
        \and
    Institute for Astronomy \& Astrophysics, Space Applications \& Remote Sensing, National Observatory of Athens, GR-15236 Palaia Penteli, Greece
        \and
    CEA Saclay, DRF/Irfu/DEDIP/LILAS, F-91191 Gif-sur-Yvette Cedex, France
        \and
    National Observatory of Athens, GR-11810 Thessio, Greece
        \and
    Physics Department, Aristotle University of Thessaloniki, GR-51124 Thessaloniki, Greece
             }

\date{}
 
\abstract
{We present the forward cosmological analysis of an $XMM$ selected sample of galaxy clusters out to a redshift of unity. Mass-observable relations have been derived in a self-consistent manner using the sample alone. Special care is given to the modelling of selection effects.} 
{Following our previous 2018 study based on the dn/dz quantity alone, we perform an upgraded cosmological analysis of the same XXL C1 cluster catalogue (178 objects), with a detailed account of the systematic errors. The results are combined with external constraints from BAO and the CMB.}
{The study follows the ASpiX methodology: the distribution of the observed X-ray properties of the cluster population is analysed in a 3D observable space (count rate, hardness ratio, redshift) and modelled as a function of cosmology along with the scaling relations and the selection function. Compared to more traditional methods, ASpiX allows the inclusion of clusters down to a few tens of photons and is of much simpler use. Two M-T relations are considered: from the CFHT and from the more recent Subaru lensing analyses.}
{We obtain an improvement by a factor of  2, compared to the previous analysis dealing with the cluster redshift distribution, for the XXL sample alone and letting the normalisation of the M-T relation and the evolution of the L-T relation free. Adding constraints from  the XXL cluster 2-point correlation function and the BAO from various surveys  decreases the uncertainties by 23 and 53 \% respectively  and 62\% when adding both. The central value is in excellent agreement with the Planck CMB constraints. Switching to the scaling relations from the Subaru analysis, and letting free more parameters, provide less stringent constraints, but still consistent with the Planck CMB at the 1-sigma level. Our final constraints are $\sigma_8$ = $0.99^{+0.14}_{-0.23}$, $\Omega_m$ = 0.296 $\pm$ 0.034 ($S_8 = 0.98^{+0.11}_{-0.21}$) for the XXL sample alone. Combining XXL ASpiX, the XXL cluster 2-point correlation function and the BAO, letting 11 free parameters and allowing for the cosmological dependence of the scaling relations in the fit, induce a shift of the central values, which is reminiscent of that observed for the Planck S-Z cluster sample. We find $\sigma_8$ = $0.793^{+0.063}_{-0.12}$, $\Omega_m$ = 0.364 $\pm$ 0.015 ($S_8 = 0.872^{+0.068}_{-0.12}$), but still compatible with Planck CMB at 2.2$\sigma$.}
{The results obtained by the ASpiX method are promising; further improvement is expected from the final XXL cosmological analysis involving a cluster sample twice as large. Such a study paves the way for the analysis of the  eROSITA and future Athena surveys.}  
 

\titlerunning{The XXL survey XLVI: Forward cosmological analysis of the C1 cluster sample}
\authorrunning{C. Garrel et al.}
\maketitle


\section{Introduction}

As the largest gravitationally collapsed objects in the universe, clusters of galaxies occupy a twofold-privileged position in astrophysical studies. The cluster number counts and spatial distribution of galaxy clusters as a function of mass and redshift is sensitive to both the growth of structure and geometry of the universe, hence constituting a powerful cosmological probe. While the purely gravitational aspect is theoretically well understood the interplay between the three cluster components, galaxies ($\sim$ 5\%), gas ($\sim$ 15\%) and dark matter ($\sim$ 80\%) renders the physics of the intracluster medium (ICM) complex. A wide range of phenomena are involved: cooling through X-ray emission, enrichment and heating of gas through supernovae and AGN feedback, turbulence and magnetic fields \citep[see e.g. review by][]{Allen2011}. These processes make clusters interesting astrophysical laboratories and have motivated considerable computational efforts to reproduce their properties using hydrodynamic simulations \citep[e.g. review by][]{Borgani2011}. The modelling of these properties is crucial in linking cluster observables like galaxy richness, velocity dispersion, gas mass, X-ray luminosity and temperature ($L_{\rm X}$, $T_{\rm X}$) to the total cluster mass - a key component for cosmological studies.\\
In this context, a Very Large XMM programme was allocated in 2010: with its two spatially disconnected regions of 25 deg$^2$ each, the XXL survey was specifically designed to obtain robust cosmological constraints from the X-ray cluster population out to a redshift of unity. It is accompanied by an extensive multi-wavelength follow-up programme and has motivated the development of sophisticated detection and analysis procedures \citep[][hereafter XXL paper I]{Pierre2016}. We refer the reader to this paper for a comprehensive bibliographical overview of cosmological X-ray cluster surveys.
In the construction of the XXL cluster sample, two aspects were given special attention.
 (i) The cluster selection is solely described in terms of observed X-ray parameters: by selecting clusters in the two-dimensional count-rate vs. apparent-size parameter space, we can ensure a sample  better than 95\% pure and whose definition is independent of the cosmology. (ii) The cluster scaling relations entering the cosmological analysis are derived from the cluster sample data alone. \\
A first cosmological analysis of the brightest XXL clusters (the C1 sample containing 178 objects) was presented in \citet[hereafter XXL paper XXV]{Pacaud2018}. This study, based on the modelling of the cluster redshift distribution (dn/dz), provided constraints on $\sigma_8$ and $\Omega_m$ with a precision of the order of 10\% and 20\% respectively. No cluster mass information was propagated in the analysis, other than the resulting mass detection limit as a function of redshift and cosmology. A natural follow-up would be the subsequent analysis of the dn/dM/dz distribution, which is theoretically much more constraining than dn/dz. However, because the direct handling of the (cosmology-dependent) masses is difficult, we adopted a forward modelling based on the prediction of directly observable quantities; namely, the three-dimensional distribution of the count-rates (CR), hardness ratios (HR) and redshifts of the selected cluster population (X-ray observable diagrams, hereafter XOD). This method (named ASpiX) has been initiated by \cite{Clerc2012a,Clerc2012b} and further validated on analytical and numerical simulations \citep{Pierre2017,Valotti2018}. ASpiX is intrinsically equivalent to the study of the mass-redshift distribution since the mass information is encoded in the CR-HR-z distribution, but it is of much simpler use and less subject to physics/cosmology degeneracies. The method consists in comparing the observed XOD with the predicted XODs as a function of cosmology and cluster evolutionary physics. In the present study, the comparison is performed adopting a Markov Chain  Monte-Carlo (MCMC) approach, in which selections of cosmological parameters and scaling relation coefficients are free; the predicted XOD are convolved by a realistic measurement error model.\\
\newline
The paper is organised as follows. Sec. 2 briefly recalls the main properties of the cluster sample. In Sec. 3, we describe the steps involved in the XOD construction. Sec. 4 performs a first cosmological analysis under exactly the same hypotheses as in XXL paper XXV; this allows a direct comparison of the two approaches. We further add constraints from the two-point correlation function from the same cluster sample. In Section 5, we actualise the study by using the revised scaling relations obtained from our recent lensing analysis of deep Hyper Suprime Camera images. The results along with various sources of uncertainty are discussed in Sec. 6, with constraints from other probes. Conclusions are drawn in Sec. 7. Appendix \ref{measurement_sec} describes the procedure used to measure the cluster quantities appearing in the XOD. Appendix \ref{like_sec_appendix} gives the details of the cosmological likelihood calculation in the CR-HR space, including the estimate of the sample variance.\\
Throughout the paper, we assume a spatially flat $\Lambda$ cold dark matter ($\Lambda$CDM) model (see Section 4). We use the standard notation $M_{\Delta}$ to denote the cluster mass enclosed within a sphere of radius $r_{\Delta}$, within which the mean overdensity equals $\Delta$ times the critical density of the universe at a particular redshift $z$.

\section{Cluster sample}\label{sec:cluster_sample}
The  present paper deals with a sample of 178 XXL  C1 clusters detected in the 47.36 $\mathrm{deg^2}$  XXL survey. It is identical to the sample used in the previous XXL cosmological analysis (XXL paper XXV).
In this section, we recall the properties of this sample and describe the measurements of the cluster parameters used in the current study. 

\subsection{Sample}

\begin{figure}
   \centering
   \includegraphics[width=0.45\textwidth]{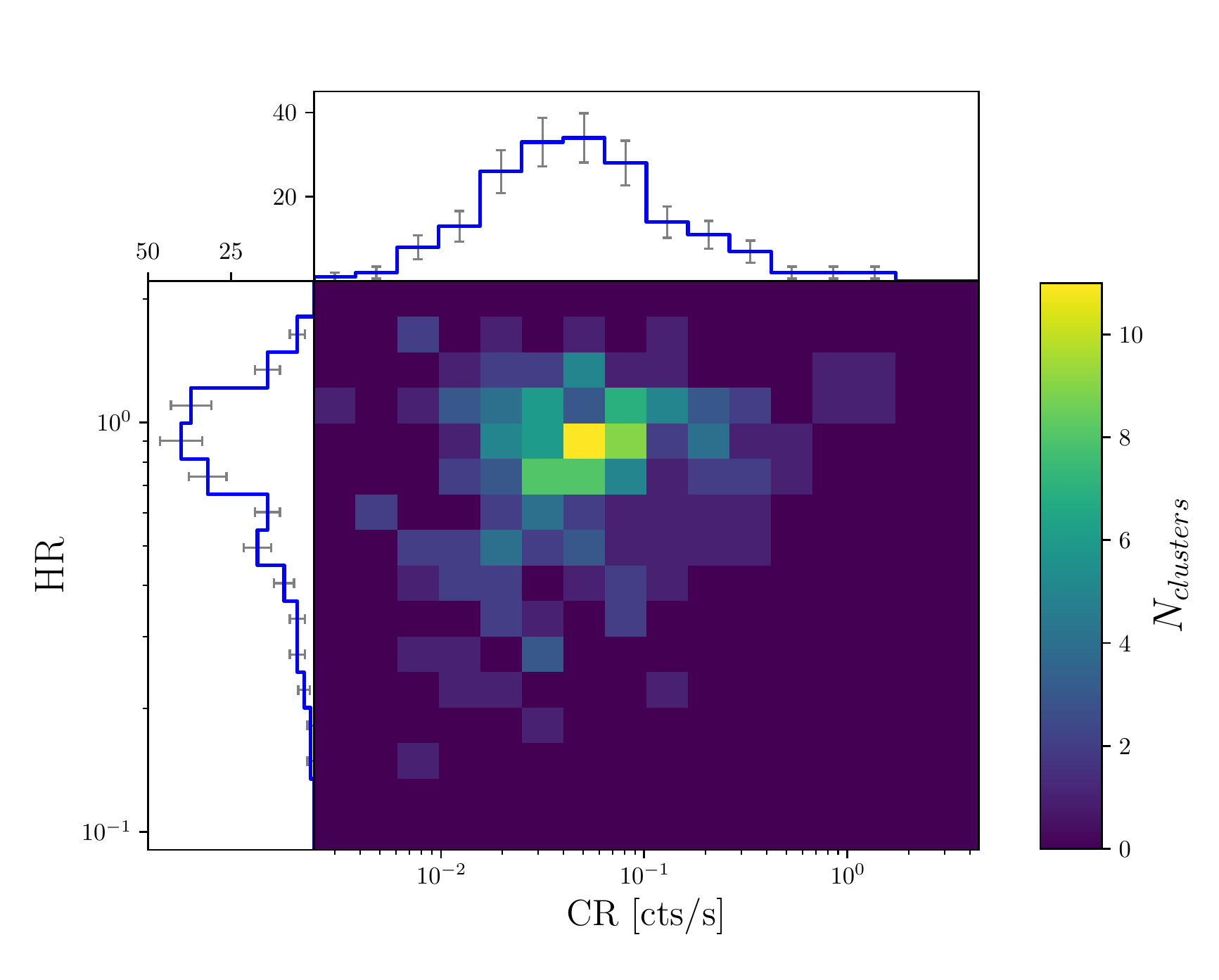} \hfill
   \caption{The X-ray observable diagram (XOD) of the XXL C1 sample containing 178 objects, integrated over the [0.05-1] redshift range. The blue histograms show the 1D integrated CR and HR distributions.  Error bars only account for shot noise.}
   \label{Diag_178}%
\end{figure}

\cite{Adami2018}, hereafter XXL paper XX, published a sample of 365 clusters divided in two classes, namely the C1 and C2 class. The C1 sub-sample consists of 191 sources, achieving a high degree of purity (fewer than 5\%  of  point sources misclassified as extended \citep{Pacaud2006}). We restricted the cosmological analysis to only the C1 spectroscopically confirmed clusters within the [0.05-1] redshift range. This led to the exclusion of 8 clusters that were outside of the redshift range and 5 without redshift estimates, resulting in a final sample of 178 clusters.
In the present study, each cluster is characterised by four observable parameters:  redshift ($z$),  X-ray count-rate (CR, defined as the number of X-ray counts received per second in the [0.5-2] keV energy band, normalised to its on-axis value), hardness ratio (HR, defined as the ratio between the count-rates in the [1-2] and [0.5-1] keV energy bands), and angular core radius \citep[$\theta_c$, assuming a $\mathrm{\beta}$-profile with $\beta$=2/3;][]{Cavaliere1976,Cavaliere1978};  CR and HR are equivalent to physical flux and colour.\\
The CR-HR distribution of the XXL C1 sample is shown in Fig.~\ref{Diag_178}.

\subsection{Measurements} \label{sec_measurement_sec}

Along with a precise mapping of the selection function, the ASpiX method requires robust measurements of the CR and HR quantities with realistic error estimates. We describe below the adopted procedure.\\
The XXL detection pipeline \citep[{\sc Xamin};][]{Pacaud2006} operates on the [0.5-2] keV images and provides a list of sources; a multi-PSF fit  returns the source extent ($\theta_{c}$) for the extended source model, and the resulting CR. The X-ray pipeline also provides the pixel segmentation mask for each source from the first detection pass. The fitted CR and $\theta_c$ values are ascribed measurement likelihoods by {\sc Xamin}, but errors are not provided for these quantities. In all what follows, the  {\sc Xamin} output is used to deal with cluster selection issues only.\\
In order to obtain model-independent CR measurements along with associated errors,  we apply a novel method based on Monte-Carlo sampling to fit  the X-ray profile \citep[{{\tt pyproffit}};][]{Eckert2020} on the mosaic of overlapped XMM observations. Based on a multi-scale profile decomposition, this method allows robust CR and, thus, HR measurements, together with an estimate of their uncertainties. Given that the mean number of collected photons per cluster is low (on average $\sim$ 200 counts for 10 ks exposures),  we  use a simplified  minimisation algorithm for the  $\theta_{c}$ measurements.  
The complete procedure is detailed in Appendix \ref{measurement_sec}.

\section{Cosmological modelling}

The main goal of the paper is to perform a forward cosmological analysis : the ASpiX method consists of fitting the $\mathrm{CR}$-$\mathrm{HR}$-$z$ XOD. This approach was tested on simulations and described in a series of articles \citep{Clerc2012a,Clerc2012b,Pierre2017,Valotti2018}. In this section, we recall the principles and assumptions inherent to the method.  

\subsection{ASpiX method} \label{aspix_meth}

Starting from a theoretical mass function, ASpiX reconstructs the XOD of a given cosmological plus cluster physics model, in order to match the observed XOD. The strength of this method is to rely only on strictly observable X-ray parameters, which means that the cluster temperatures, luminosities and masses are not explicitly computed.\\
We start from the differential mass function computed for a given cosmology, and expressed in terms of redshift ($z$) and sky area ($\mathrm{\Omega}$) folded with the XXL survey effective sky-coverage. We compute the distribution in terms of $M_{\Delta}$, $z$ and of the cluster characteristic size:  
\begin{equation}
r_{\Delta} = \left(\frac{3M_{\Delta}}{4\pi\times\Delta\rho_c}\right)^{1/3}
\end{equation}
where $\mathrm{r_{\Delta}}$ is the radius within which the average density is $\mathrm{\Delta}$ times $\mathrm{\rho_c}$, the critical density.\\
We use scaling relations linking mass and temperature [$\mathrm{M_{\Delta}}$-T], luminosity and temperature [L-T] and between $\mathrm{r_{\Delta}}$ and the cluster core radius $\mathrm{r_c}$, assuming a $\beta$ model with $\beta$=2/3,  [$\mathrm{r_c}$-$\mathrm{r_{\Delta}}$].   The relation between physical and apparent core radii reads: 
\begin{equation}
\theta_c \mathrm{\ [arcsec]} = \frac{648000}{\pi} r_c \mathrm{\ [Mpc]}\ /\ d_{a}(z) \mathrm{\ [Mpc]} 
\end{equation}
where $d_a$ is the angular diameter distance. \\
We make use of the {\sc apec} model: the emission spectrum from collisionally-ionised diffuse gas is calculated from the {\sc AtomDB} atomic database, \citep{Smith2001} with a metallicity of 0.3 $\mathrm{Z_\odot}$. Folding the spectrum with  the EPIC (European Photon Imaging Camera) $XMM$ response matrices, provides us with  count-rates in the three energy bands of interest ([0.5-2] keV, [0.5-1] keV and [1-2] keV). From this, we subsequently construct the 4D $z$-CR-HR-$\mathrm{\theta_c}$ diagram.\\
We stress  that the mass information is implicitly encrypted in the $\theta_c$, CR and HR parameters via the scaling relations. We then apply the XXL survey selection function (f[CR,$\mathrm{\theta_c}$]) and, finally, convolve the XOD with the measurement-error model of each observable parameter except for $z$ (spectroscopically measured and thus with negligible error).\\
In the end, we integrate over $\mathrm{\theta_c}$ to obtain the 3D $z$-CR-HR diagram expected for a given cosmology.

\subsection{Assumptions and Numerical Inputs} \label{sec:numerical_inputs}

For the purpose of this analysis, we use a Tinker mass function \citep{Tinker2008} computed at an overdensity of $\mathrm{\Delta}$ = 500. We disperse over the luminosity, temperature and core radius distributions by including a log-normal scatter around the mean scaling relations. The binning of the XOD is shown in Table~\ref{inputXOD}.
\begin{table}
\captionsetup{singlelinecheck = false, justification=justified}
     \caption{Sampling of the X-ray parameter distribution in the XOD.}
     $$ 
         \begin{tabular}{lccc}
            \hline
            \noalign{\smallskip}
            Parameter      &  Min, Max &    Nb. of bins      &  Scale\\
            \noalign{\smallskip}
            \hline
            \noalign{\smallskip}
            $z$ & 0.05, 1 & 5 & linear\\
            CR [cts/s] & 0.002, 4.4 & 16 & log\\
            HR & 0.09, 2.2 & 16 & log\\
            $\theta_c$ $\rm{[arcsec]}^{\ *}$ & 3, 211 & 16 & log\\
            \noalign{\smallskip}
            \hline
         \end{tabular}
     $$ 
      \tablefoot{ $\mathrm{^{(*)}}$ The selection function is applied in the CR-$\theta_c$ plane, but the $\theta_c$ distribution is not directly used to constrain the cosmological parameters. The  XOD is then  integrated over the $\mathrm{\theta_c}$ and convolved with the error model,  prior to the fit.}
         \label{inputXOD}
\end{table}

\subsubsection{Scaling relations}\label{scal_rel_sec}

\begin{table*}
\captionsetup{singlelinecheck = false, justification=justified}
     \caption{Cluster scaling laws used in the first part  of the paper.}
     $$ 
         \begin{tabular}{lcccc}
            \hline
            \noalign{\smallskip}
            Law      &  $\mathrm{X_0}$      &  $\mathrm{\alpha}$      &  $\mathrm{\gamma}$ & Scatter\\
            & & & & (natural log)\\
            \noalign{\smallskip}
            \hline
            \noalign{\smallskip}
            $M_{500,WL}$-$T_{300kpc}$ & $(2.6 \pm 0.55)\times 10^{13} \mathrm{M_{\odot}\cdot h^{-1}}$ & 1.67 & -1.0 & -\\
            $L^{XXL}_{500,WL}$-$T_{300kpc}$ & $8.24\times 10^{41} \mathrm{erg\cdot s^{-1}}$ & 3.17 & $\mathrm{0.47 \pm 0.68}$ & 0.67\\
            $\mathrm{r_c}$-$\mathrm{r_{500}}$ & 0.15 & - & - & -\\
            \noalign{\smallskip}
            \hline
         \end{tabular}
     $$ 
      \tablefoot{To keep the same configuration as in XXL paper XXV, we include scatter only in the luminosity distribution. We use a log-normal scatter, with value indicated in the table, as in the previous XXL cosmological analysis (XXL paper XXV). Uncertainties on parameters indicate that these parameters are kept free during the analysis, in a Gaussian prior range with $\sigma$ given by the uncertainties and then marginalised over. The mass-temperature relation was published in the first XXL release \citep[][hereafter XXL paper IV]{Lieu2016} and the luminosity-temperature relation, in the second XXL release, XXL paper XX.}
         \label{tab_scalrel}
\end{table*}

In the first part of the paper, in order to allow a direct comparison of the different methodologies applied, we stick to the scaling relations of XXL paper XXV  modelled, as usual, by power laws:
\begin{equation}
\frac{M_{500,WL}}{X_{0,M-T}} = \left(\frac{T_{300kpc}}{1~{\rm keV}}\right)^{\alpha_{M-T}} E(z)^{\gamma_{M-T}}
\end{equation}
\begin{equation}
\frac{L^{XXL}_{500,WL}}{X_{0,L-T}} = \left(\frac{T_{300kpc}}{1~{\rm keV}}\right)^{\alpha_{L-T}} E(z)^{\gamma_{L-T}}
\end{equation}
\begin{equation}\label{eq_scal_rc}
\frac{r_c}{X_{0,r_c-r_{500}}} = r_{500}
\end{equation}
where $\mathrm{M_{500,WL}}$ is the weak lensing mass estimate within $r_{500}$, $\mathrm{T_{300kpc}}$ the cluster X-ray temperature measured inside 300 kpc, $\mathrm{L^{XXL}_{500,WL}}$ the luminosity within $r_{500}$ in the [0.5-2] keV energy band and $E(z)$ is the redshift evolution of the Hubble parameter, $E(z) \equiv H(z)/H_0$. The mass calibration  only relies on weak lensing measurements based on the CFHT lensing data \citep[for a didactic review of cluster weak lensing, see][]{Umetsu2020b}. The scaling law parameters are summarised in Table~\ref{tab_scalrel}.\\
During the analysis, two scaling relation parameters were introduced as nuisance parameters and marginalised. These parameters correspond to the ones indicated by uncertainties in Table~\ref{tab_scalrel}. We do not include any scatter in the $r_c-r_{500}$ and $M_{500,WL}$-$T_{300kpc}$ relations within the base model for the purposes of comparing to XXL paper XXV. Subsequently, we add a scatter of 0.1 in the $r_c-r_{500}$ in Section \ref{sec:mauro_laws} to correspond to the actualised scaling relations. We discuss the impact of larger values of the scatter in Section \ref{subsection_discuss_sr}.

\subsubsection{Selection function}

\begin{figure}
   \centering
   \includegraphics[width=0.45\textwidth]{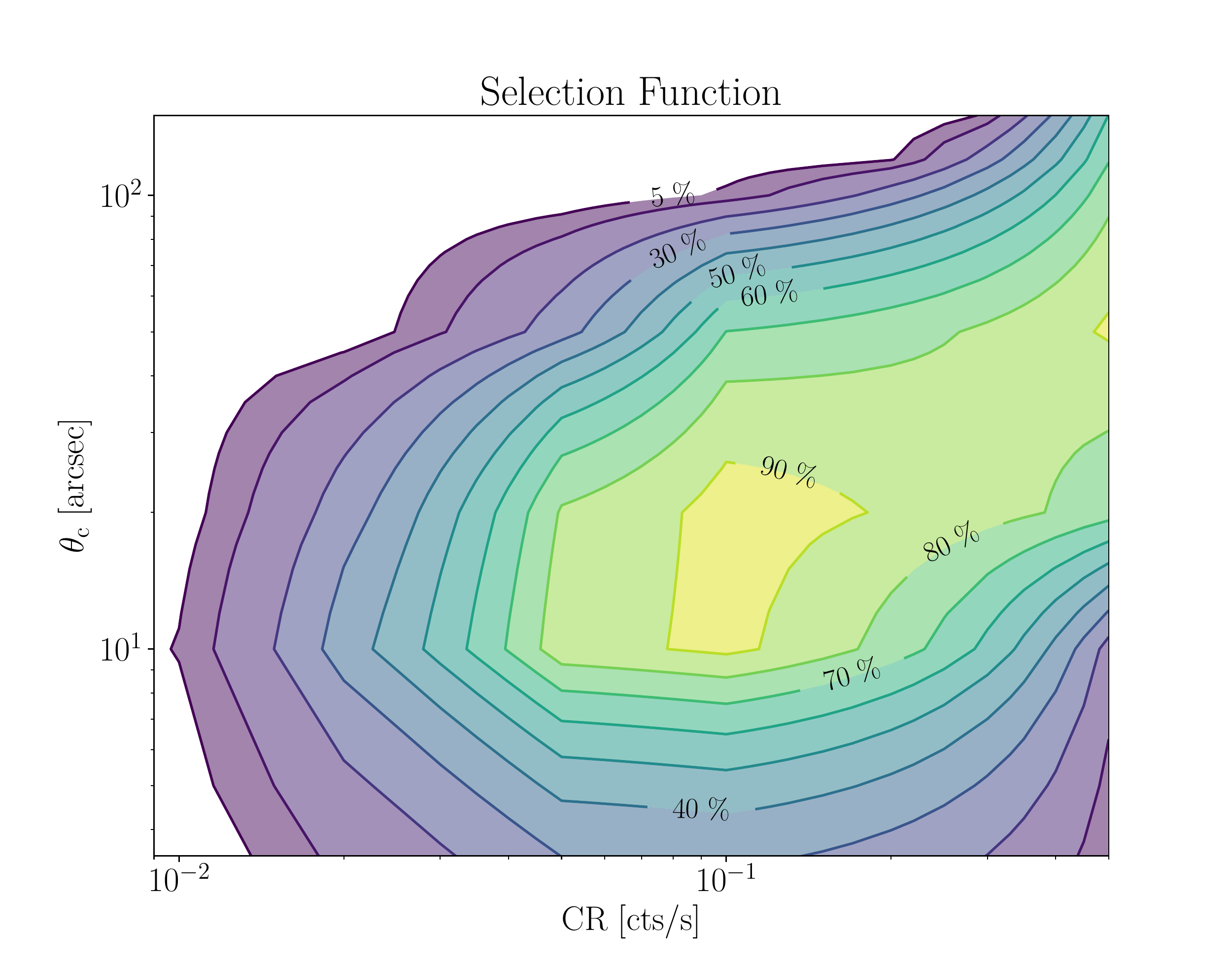} \hfill
   \caption{The XXL C1 selection function used in this analysis. From simulations of $XMM$ cluster observations, the detection probability is expressed as a function of only observable quantities: the count rate and the apparent size of a $\beta$=2/3 model, $\theta_c$. The same selection function was used in XXL paper XXV.}
              \label{selfunc}%
    \end{figure}
Assuming a circular $\beta$=2/3 model for extended sources, a cluster population with different count-rates and $\theta_c$, was generated for a range of  $XMM$ exposures. This process takes into account the instrumental characteristics (PSF distortion, vignetting, detector masks, background and Poisson noise) for the three $XMM$ detectors. Point sources are added at random over the $XMM$ field of view, with a flux distribution following the $\log(N)$-$\log(S)$ from \cite{Moretti2003} down to $\mathrm{5\times10^{-16}\ erg\ s^{-1}\ cm^{-2}}$; the contribution of point sources below $\mathrm{4\times10^{-15}\ erg\ s^{-1}\ cm^{-2}}$ is included in the cosmic background component \citep{Read2003}. The XXL cluster detection algorithm is then applied, allowing a statistical study to determine various levels of completeness and purity. The cluster selection  is performed in the {\sc Xamin} output parameter space ({\sc ext, ext\_stat}) and subsequently translated into the CR-$\theta_c$ plane. The details of the procedure are given in \cite{Pacaud2006}. The XXL C1 selection function, matched to the XXL exposure and background maps, is shown in Fig.~\ref{selfunc}.\\
We stress that the selection function is mapped back into the intrinsic CR-$\theta_c$ space (the probability to detect a cluster that has these parameters - not the pipeline ones measured at those values); there is therefore no inconsistency in using for the cosmological analysis, CR values that were measured using the {\tt pyproffit} package.

\subsubsection{Measurement errors} \label{classic_errmod_sec}

\begin{table}
\captionsetup{singlelinecheck = false, justification=justified}
\caption{The values of the $\{a_x\}$, $\{b_x\}$ and $\{c_x\}$ coefficients from equation \ref{errmod_func}.}
     $$ 
         \begin{tabular}{lccc}
            \hline
            \noalign{\smallskip}
            Error model      &  $a_x$ &    $b_x$      &  $c_x$\\
            \noalign{\smallskip}
            \hline
            \noalign{\smallskip}
            $\%err_{CR}$ & 0.0155 & -0.488 &  0.112\\
            $\%err_{HR}$ & 0.0298 & -0.488 &  0.143\\
            $\%err_{\theta_c}$ & 0.0567 & -0.432 & -0.133\\
            \noalign{\smallskip}
            \hline
         \end{tabular}
     $$ 
      \tablefoot{The functions, $\%err_x(\mathrm{CR},\theta_c)=a_x \ \mathrm{CR}^{b_x} \ \theta_{c}^{\ c_x}$, are fitted with the data using the Levenberg-Marquardt algorithm.}
         \label{errmod_coef}
\end{table}
\begin{figure*}
   \centering
   \includegraphics[width=0.33\textwidth]{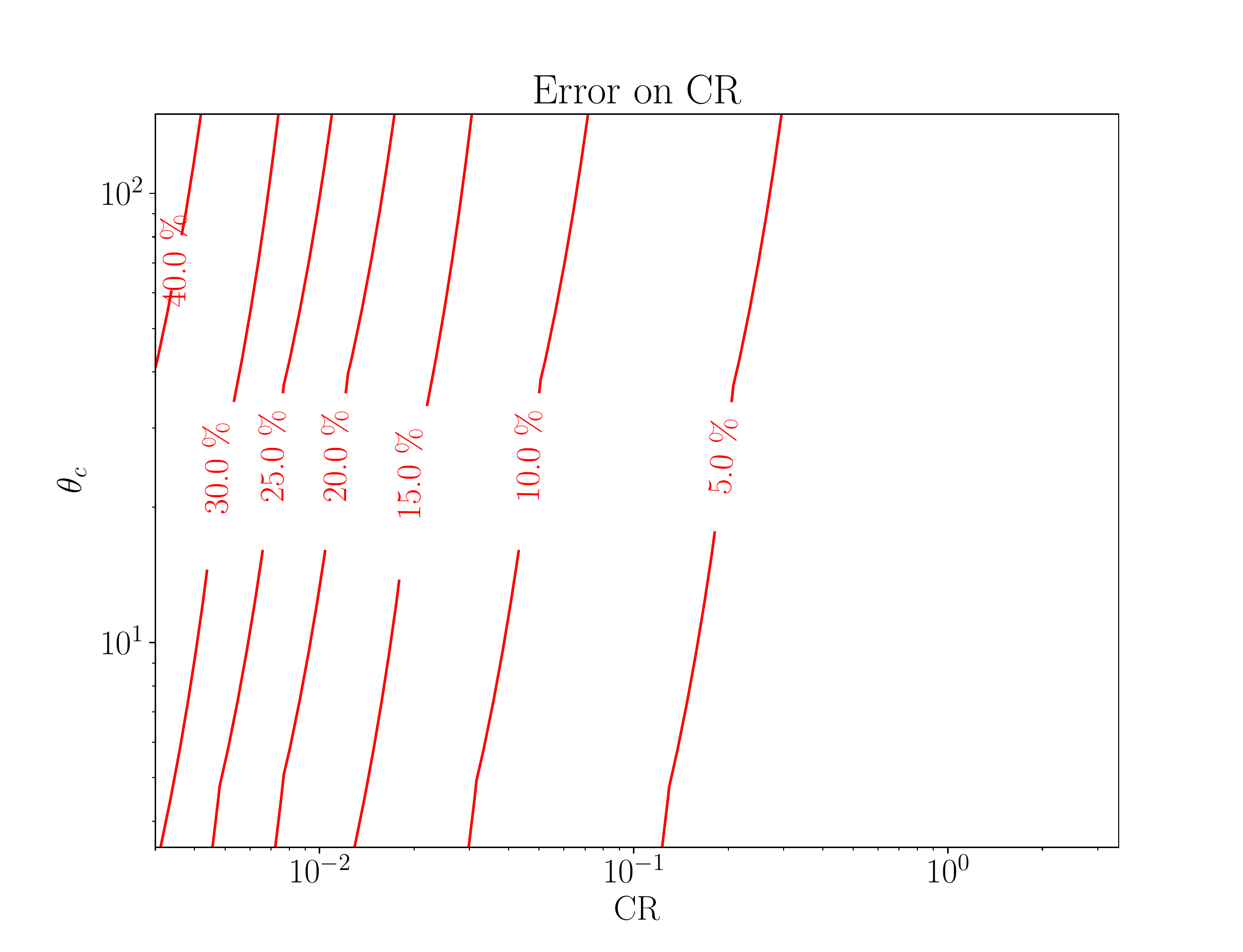} \hfill
   \includegraphics[width=0.33\textwidth]{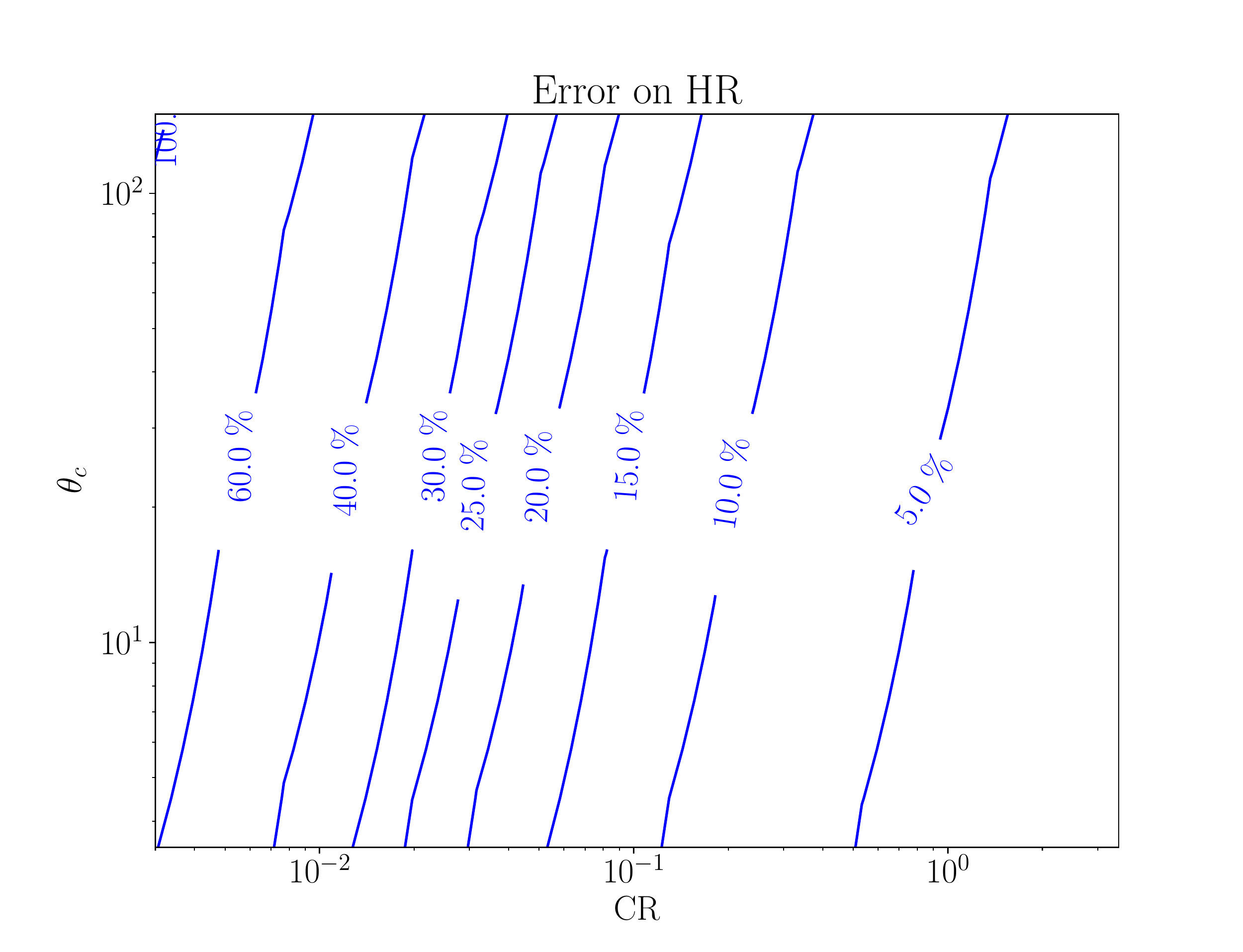} \hfill
   \includegraphics[width=0.33\textwidth]{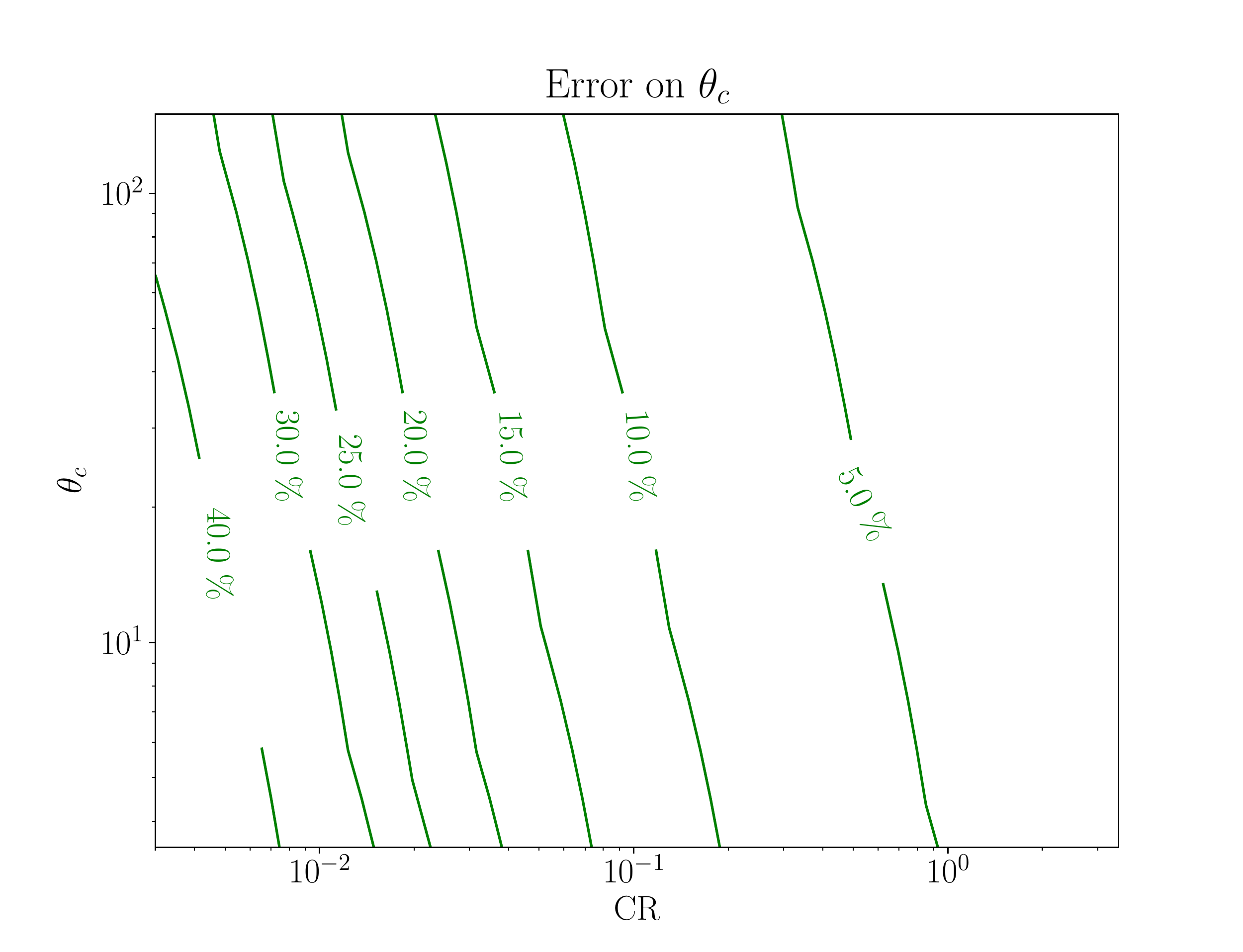} \hfill
   \caption{The relative error model (in percent) for each observable used in this analysis. These models are computed as a function of CR and $\theta_c$.}
              \label{errmod}%
\end{figure*}
 As shown in \cite{Clerc2012b}, the inclusion of measurement errors changes the shape of the predicted X-ray observable diagrams. A precise estimate of the measurement errors is thus a key step  of the analysis.\\
The {\tt pyproffit} package provides us with error estimates for each measurement. This allows us to subsequently model the relative measurement errors on  CR, HR  and $\theta_c$  as a function of CR and $\theta_c$, using the following parametrisation: 
\begin{equation}\label{errmod_func}
\%err_{x} = a_x \ \mathrm{CR}^{b_x} \ \theta_{c}^{\ c_x} ~~~~ (x = CR, HR, \theta_c )
\end{equation} 
The choice of  the CR-$\theta_c$ plane is a natural second order approximation, reflecting the fact that, physically, the brighter and more peaked a cluster, the better the measurement. \\
We perform a non-linear least square fit using the Levenberg-Marquardt algorithm \citep{Levenberg,Marquardt} to constrain the $\{a_x\}$, $\{b_x\}$ and $\{c_x\}$ coefficients. The resulting error models are shown in Fig.~\ref{errmod} and the coefficients are given in Table~\ref{errmod_coef}.

\subsubsection{Likelihood}\label{like_sec}

The log-likelihood model to infer the cosmological parameters is given, for each redshift bin, by\footnote{See Appendix \ref{like_sec_appendix} for its formal derivation}:
\begin{equation}
\begin{aligned}
{\cal L}_{z_i} = & \bar n - \sum_j \hat N_j \ln(\bar n_j)  \\
& + \frac{1}{2} \mathrm{ln}\left(1 + \sigma^2_{\delta} \sum_j \hat N_j \frac{\bar b_{j}}{\bar b} \right)\\
& - \frac{\sigma^2_{\delta}}{2} \times \left( \sum_j \hat N_j \frac{\bar b_{j}}{\bar b} 
- \bar n \right)^2 \left( 1+\sigma^2_{\delta} \sum_j \hat N_j 
\frac{\bar b^2_{j}}{\bar b^2} \right)^{-1}  \hspace{0.5cm} 
\label{eq:lnLzi-2_sec}
\end{aligned}
\end{equation}
where $\bar n_j$ is the number of predicted clusters in the ($\mathrm{CR}_j$,$\mathrm{HR}_j$) 2D bin (and $\bar n$ the number of predicted clusters in the redshift bin $i$) and $\hat N_j$ is the number of observed clusters. $\bar b_j$ is the mean galaxy cluster bias for the 2D bin $j$ and $\bar b$ the mean bias of the survey (see Eq.~\ref{eq:b-i-def}
 and \ref{eq:N-tot-def} of Appendix \ref{like_sec_appendix}). To calculate these quantities, we use the \cite{Tinker2010} bias model. $\sigma^2_{\delta}$ is the variance of the total number density contrast. Then, in Eq.~\ref{eq:lnLzi-2_sec}, the first line is the usual shot-noise term and the second and third lines are the sample-variance terms.\\
Finally:
\begin{equation}
{\cal L} = \sum_i {\cal L}_{z_i}
\end{equation}
Following the formalism presented in \cite{Valageas2011}, we estimate that the sample variance value is $\mathrm{\sim}$30\% of the Poisson variance.\\
\newline
The cosmological parameters are constrained using a Markov Chain Monte-Carlo procedure by an affine-invariant ensemble \citep{Goodman2010}, following the {\tt EMCEE} algorithm \citep{Foreman-Mackey2013}. In order to optimise computational time while having enough statistics to control the chain convergence, we ran five independent chains in parallel, each chain had 2$N$ walkers, with $N$ specifying the number of free parameters. The chains are stopped when reaching the Gelman-Rubin convergence criterion of R-1 $<$ 0.03, after excluding a 20$\%$ burn-in phase.

\section{Cosmological analysis with the scaling relations of XXL paper XXV}

We assume a flat $\Lambda \mathrm{CDM}$ model. We perform the Monte Carlo analysis and create contour plots by means of the {\sc getdist} python package \citep{Lewis2019}.\\
The displayed 1$\sigma$ and 2$\sigma$ confidence intervals  show respectively the 68\% and 95\% limits. 

\subsection{XXL ASpiX alone} \label{main_results}

\begin{table*}
\captionsetup{singlelinecheck = false, justification=justified}
\caption{ASpiX cosmological constraints for the base model and the joint analysis (flat $\Lambda$CDM).}
     $$ 
         \begin{tabular}{*{6}{P{2.3cm}}}
            \hline
            \noalign{\smallskip}
            \tiny Parameter & \tiny XXL ASpiX & \tiny XXL ASpiX & \tiny XXL ASpiX & \tiny XXL ASpiX & \tiny Priors\\
             & \tiny Base & \tiny + XXL clustering & \tiny + XXL clustering & \tiny + Planck CMB &\\
              &  &  & \tiny + BAO & &\\
            \noalign{\smallskip}
            \hline
            \noalign{\smallskip}
            $\Omega_m$\dotfill&\small $0.342^{+0.038}_{-0.046}$ &\small $0.314 \pm 0.031$ &\small $0.317 \pm 0.017$ &\small $0.317 \pm 0.007 $ & \small $\mathcal{U}(0.09,1.0)$\\
            \noalign{\smallskip}
            $\sigma_8$\dotfill &\small $0.829 \pm 0.048$ &\small $0.840 \pm 0.044$ &\small $0.838^{+0.035}_{-0.042}$ &\small $0.811 \pm 0.006$ & \small $\mathcal{U}(0.05,2.0)$\\
            \noalign{\smallskip}
            $S_8$\dotfill &\small $0.882 \pm 0.046$ &\small $0.857^{+0.042}_{-0.050}$ &\small $0.861^{+0.033}_{-0.042}$ &\small $0.834 \pm 0.011$ & -\\
            \noalign{\smallskip}
            $\Omega_b$\dotfill &\small $0.049 \pm 0.004$ &\small $0.046 \pm 0.001$ &\small $0.046 \pm 0.001$ &\small $0.0495 \pm 0.0006$ & \small $\mathcal{N}(0.0493,0.0035^2)$\\
            \noalign{\smallskip}
            $h$\dotfill & - &\small $0.638^{+0.014}_{-0.035}$ &\small $0.627^{+0.011}_{-0.018}$ &\small $0.672 \pm 0.005$ & \small $\mathcal{U}(0.55,0.9)$\\
            \noalign{\smallskip}
            $n_s$\dotfill &\small $0.963^{+0.021}_{-0.017}$ &\small $0.966 \pm 0.009$ &\small $0.965 \pm 0.009$ &\small $0.964 \pm 0.004$ & \small $\mathcal{N}(0.9649,0.022^2)$\\
            \noalign{\smallskip}
            $\tau$\dotfill & - & - & - &\small $0.053 \pm 0.008$ & \small $\mathcal{U}(0.01,0.1)$\\
            \noalign{\smallskip}
            $X_{0,M-T}$\dotfill & - & - & - & - & \small $\mathcal{N}(2.6,0.55^2)$\\
            \noalign{\smallskip}
            $\gamma_{L-T}$\dotfill & - & - & - & - & \small $\mathcal{N}(0.47,0.68^2)$\\
            \noalign{\smallskip}
            \hline
         \end{tabular}
     $$ 
      \tablefoot{For the base model, we do not quote constraints on $h$ since this parameter is poorly constrained by cluster counts and the posterior distributions are driven by the hard prior. $S_8$ is defined to be $S_8 \equiv \sigma_8(\Omega_m / 0.3)^{0.5}$. The last column indicates the priors used in this analysis. $\mathcal{N}(\mu,\sigma^2)$ corresponds to a Gaussian prior with mean $\mu$ and variance $\sigma^2$ and $\mathcal{U}(A,B)$ a uniform prior within the range [A,B]. We do not quote constraints on the nuisance parameter used in the analysis.}
         \label{best_fit_cosmo_wfix}
\end{table*}
\begin{figure}
   \centering
   \includegraphics[width=0.45\textwidth]{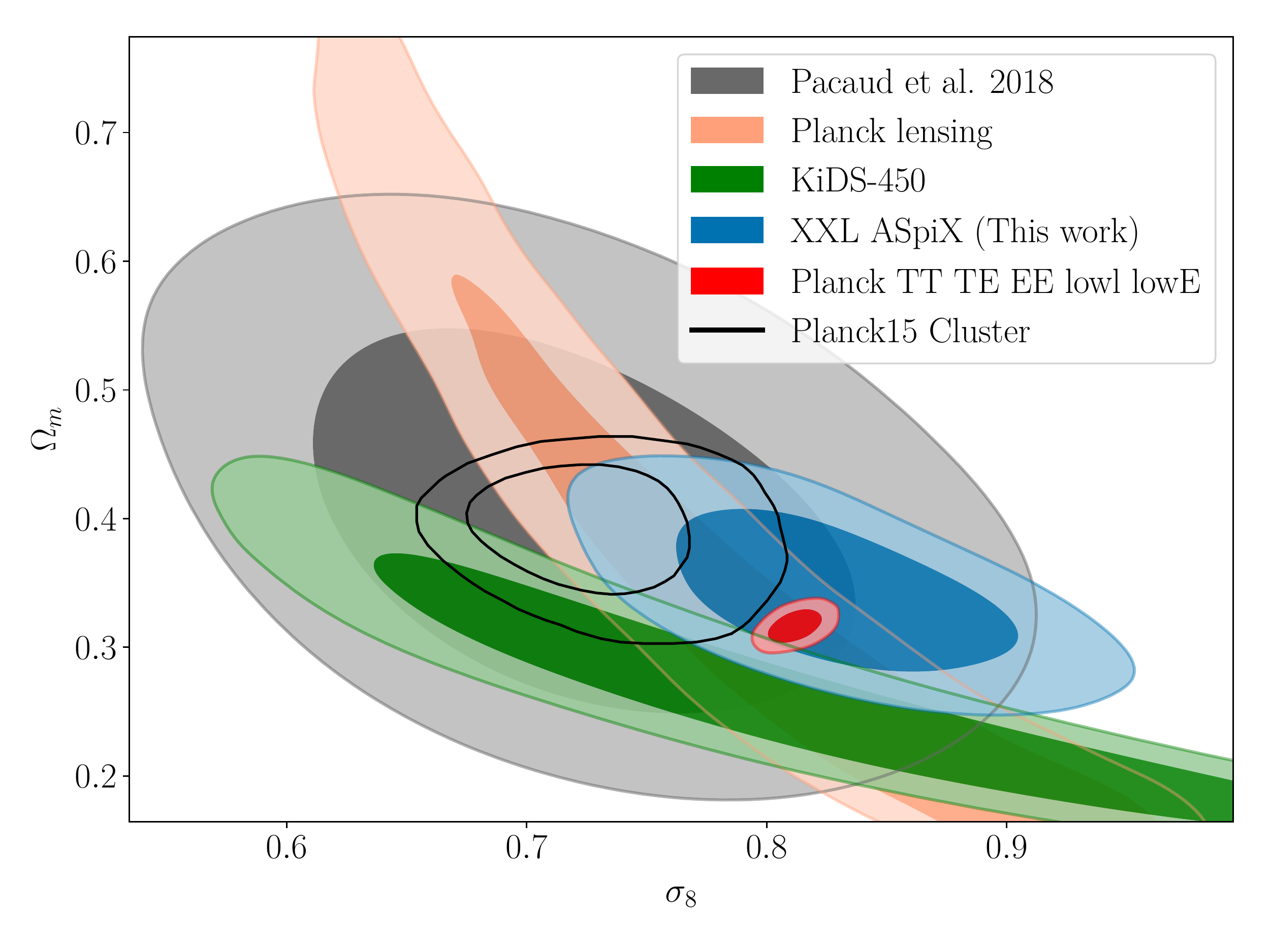} \hfill
   \caption{$\Omega_m-\sigma_8$ cosmological constraints from XXL ASpiX (this work), Planck-2018 (Planck TT TE EE lowl lowE), Planck S-Z cluster counts (Planck15 Cluster), Planck lensing \citep[CMB lensing potential analysis;][]{Planck2018b}, KIDS-450 \citep[tomographic weak gravitational lensing of the 450 deg$^2$ Kilo Degree Survey;][]{Hildebrandt2017} and the previous XXL cosmological results, XXL paper XXV \citep{Pacaud2018}}
              \label{fixw_sep}%
\end{figure}
In this section, we present the cosmological constraints obtained from the XOD alone.\\
We consider five free cosmological parameters within the $\Lambda$CDM framework: \{$\Omega_m,\ \sigma_8,\ \Omega_b,\ n_s,\ h$\}. Two scaling relation parameters are included as nuisance parameters: the $M-T$ normalisation ($X_{0,M-T}$), and the $L-T$ evolution index ($\gamma_{L-T}$) as summarised in Table~\ref{tab_scalrel}; this already allows for significant freedom in the parametrisation of cluster physics unknowns and related cosmological dependence. These two parameters are marginalised  during the Monte Carlo analysis.\\
We apply conservative Gaussian priors for the cosmological parameters which are not well constrained by cluster counts (namely $\Omega_b$ and $n_s$). They are centred on the Planck-2018 \citep{Planck2018a} values with errors multiplied by a factor of 5 in order not to force the agreement  between our results and the Planck ones, i.e.: $\Omega_b=0.0493\pm0.0035$, $n_s= 0.9649\pm0.022$. The prior on the Hubble constant is chosen to be   uniform within a [0.55-0.9] range.\\
\newline
While keeping the same sample and scaling relations as in XXL paper XXV, the ASpiX method allows us to improve constraints in the $\Omega_{m}-\sigma_{8}$ plane by  of a factor $\sim$ 2.   This improvement\footnote{We choose two different methods to estimate the improvement (or deterioration) in the parameter constraints in the $\Omega_m-\sigma_8$ plane:\\
We first use Green's theorem to compute the area, by line integral, inside the 1$\sigma$ contours in the $\Omega_m-\sigma_8$ plane for two different probes. Taking the square root of the ratio of these two areas gives us the estimated gain or loss in constraining power in the $\Omega_m-\sigma_8$ plane.\\
We also quantify this improvement using a figure of merit (FoM) defined as $\mathrm{FoM_{\Omega_m-\sigma_8}} = \sqrt{\mathrm{Cov^{-1}_{\Omega_m-\sigma_8}}}$ where $\mathrm{Cov_{\Omega_m-\sigma_8}}$ denotes the covariance matrix.} was not unexpected since the [$\mathrm{CR},~ \mathrm{HR}, ~z$] combination is comparable to the mass information, which did not enter the first XXL cosmological analysis, based only on $dn/dz$.\\ 
We note that, even though the XXL paper XXV results, Planck constraints and our new constraints are all compatible at the 2$\sigma$ level ($0.7\sigma$ posterior agreement\footnote{To compute the agreement between two different probes, we rely on the following process. We first draw a representative sample for each posterior of interest from the two probes. We compute the distance between each pair of points of these samples. We build, from this distance sample, the probability distribution using kernel density estimate. We then estimate the probability to exceed (PTE) by integrating the probability distribution over the interval [0-$P(0)$], with $P(0)$ the probability of a distance equal to 0. The same formalism is used in \cite{Bocquet2019}. The corresponding significance level is computed assuming Gaussian statistics. To insure that the results are not impacted by randomisation effects, we repeat this process one hundred times and present the mean significance level.}  in the $\Omega_m-\sigma_8$ plane between XXL ASpiX and XXL paper XXV) our central values now show a better agreement with the Planck results. We find $\Omega_m=0.342^{+0.038}_{-0.046}$ versus $0.3165 \pm 0.0085$ for Planck, $\sigma_8=0.829 \pm 0.048$ versus $0.8119 \pm 0.0074$ and $S_8=0.882 \pm 0.046$ versus $0.834 \pm 0.016$ for Planck, leading to a $0.4\sigma$ posterior agreement in the $\Omega_m-\sigma_8$ plane.\\
The results are summarised in Table~\ref{best_fit_cosmo_wfix}. The $\Omega_m-\sigma_8$ contours are shown in Fig.~\ref{fixw_sep} along with recent constraints from other cosmological probes.

\subsection{XXL ASpiX + XXL clustering}

\begin{figure}
   \centering
   \includegraphics[width=0.45\textwidth]{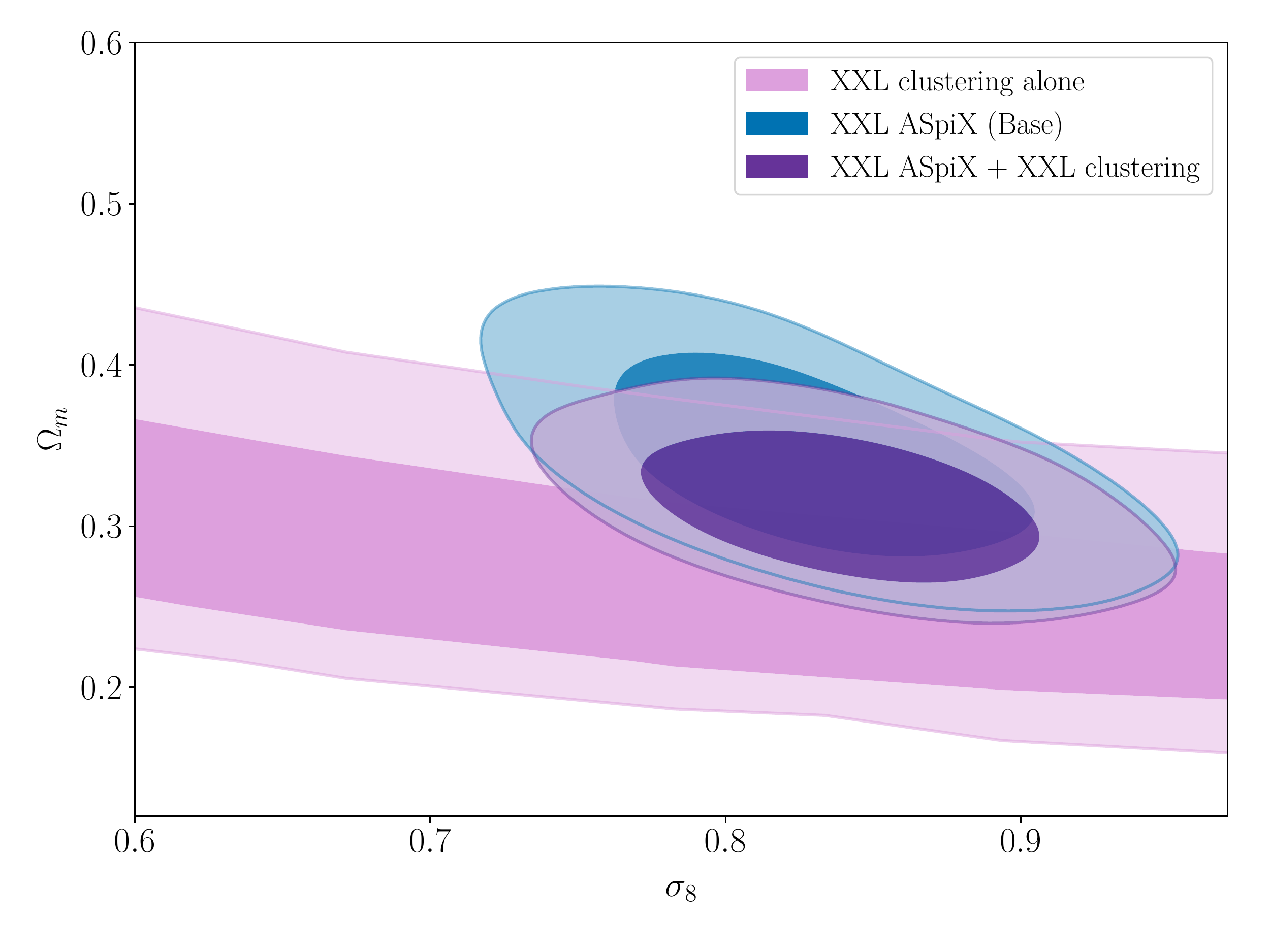} \hfill
   \caption{The $\Omega_m-\sigma_8$ cosmological constraints of XXL ASpiX cluster counts alone (base), XXL 2PCF (clustering alone) and of the joint analysis.}
              \label{clustering_aspix_joint}%
\end{figure}
Cosmological constraints from the 3D clustering analysis of the XXL cluster sample  (two-point correlation function, 2PCF), were presented in \cite{Marulli2018}.  In this section, we combine the 2PCF and  ASpiX results.\\
In order to perform the joint analysis, we run the MCMC procedure as previously, and use the XXL 2PCF mean and covariance results as additional priors for all the parameters that are left free during the analysis, namely : \{$\Omega_m,\ \sigma_8,\ \Omega_b,\ n_s,\ h,\ \tau$\}. The 2PCF study was performed using seven free parameters, the six above-mentioned cosmological parameters  and, in addition, the effective bias of the cluster sample, $b_{eff}$ \citep[see][for the detailed procedure]{Marulli2018}.\\
We model the (redshift dependent) effective sample bias following  \cite{Matarrese1997} :
\begin{equation}
    b_{eff}(z) = \mathcal{N}^{-1}(z) \int_{\mathcal{M}} \mathrm{d\ ln} M^{\prime}\ b(M^{\prime},z)\ \mathcal{N}(z,M^{\prime})
\end{equation}
and we define the averaged effective bias of the sample as:
\begin{equation}\label{beff_eq}
    b_{eff} = \sqrt{\mathcal{N}^{-2} \int_{\mathcal{Z}} \mathrm{d\ } z^{\prime}\ b_{eff}^2(z^{\prime})\ \mathcal{N}^{2}(z^{\prime})}
\end{equation}
where $\mathcal{N}(z,M)$ is the number of clusters with mass $M$ and redshift $z$ as predicted by a given cosmological scenario (including the selection effects) and $b(M,z)$, the dark matter halos bias computed using the \cite{Tinker2010} model for $M_{500}$.\\
While $b_{eff}$  is free during the XXL clustering analysis, it is an output of the cluster counts (or ASpiX) analysis. It depends on the selection function in the $M-z$ space. The selection function in the $M-z$ space depends, in turn, on the cosmology and on the scaling relations. We therefore implement the results for $b_{eff}$ from the XXL clustering analysis as an additional Gaussian term  in the likelihood from equation \ref{eq:lnLzi-2_sec}.\\
\newline
The results are shown in Table~\ref{best_fit_cosmo_wfix}. The $\Omega_m-\sigma_8$ contours are shown in Fig.~\ref{clustering_aspix_joint}. The joint XXL ASpiX + 2PCF analysis reduces the uncertainties on $\Omega_m$ and $\sigma_8$ by 23\% (FoM increased by a factor of 1.3) compared to ASpiX alone. The $\Omega_m$ result is slightly lower, $\Omega_m$ = $0.314 \pm 0.031$, and $\sigma_8$ slightly higher, $\sigma_8$ = $0.840 \pm 0.044$ and with $S_8=0.857^{+0.042}_{-0.050}$. Nevertheless, the results are still in good agreement with Planck CMB.

\subsection{XXL clusters + BAO joint analysis}

\begin{table}
\captionsetup{singlelinecheck = false, justification=justified}
\caption{BAO data used in this analysis.}
     $$ 
         \begin{tabular}{lccccc}
            \hline
            \hline
            \noalign{\smallskip}
            Survey set I & $z$ & $\mathcal{D}_{V}(z)$ & $\sigma_{\mathcal{D}}$ & $r_{s}^{fid}$ \\
            \noalign{\smallskip}
            \hline
            \noalign{\smallskip}
            \hline
            \noalign{\smallskip}
            6dFGS & 0.106 & 457 & 27 & 153.55\\
            \noalign{\smallskip}
            \hline
            \noalign{\smallskip}
            SDSS-LRG & 0.35 & 1356 & 25 & 152.76\\
            \noalign{\smallskip}
            \hline
            \noalign{\smallskip}
            SDSS-MGS & 0.15 & 664 & 25 & 148.69\\
            \noalign{\smallskip}
            \hline
            \noalign{\smallskip}
            BOSS-DR12 & 0.38 & 1477 & 16 & 147.78\\
            \noalign{\smallskip}
            \hline
            \noalign{\smallskip}
            BOSS-DR12 & 0.51 & 1877 & 19 & 147.78\\
            \noalign{\smallskip}
            \hline
            \noalign{\smallskip}
            BOSS-DR12 & 0.61 & 2140 & 22 & 147.78\\
            \noalign{\smallskip}
            \hline
             & & & & \\
            \noalign{\smallskip}
            \hline
            \hline
            \noalign{\smallskip}
            Survey set II & $z$ & $\mathcal{D}_{V}(z)$ & $\sigma_{\mathcal{D}}$ & $r_{s}^{fid}$ \\
            \noalign{\smallskip}
            \hline
            \noalign{\smallskip}
            \hline
            \noalign{\smallskip}
            WiggleZ & 0.44 & 1716 & - & 148.6\\
            \noalign{\smallskip}
            \hline
            \noalign{\smallskip}
            WiggleZ & 0.60 & 2221 & - & 148.6\\
            \noalign{\smallskip}
            \hline
            \noalign{\smallskip}
            WiggleZ & 0.73 & 2516 & - & 148.6\\
            \noalign{\smallskip}
            \hline
         \end{tabular}
     $$ 
      \tablefoot{The data come from seven different surveys: the 6dF Galaxy Survey (6dFGS) \citep{Beutler2011}, the Sloan Digital Sky Survey luminous red galaxy (SDSS-LRG) \citep{Padmanabhan2012}, SDSS data release 7 main galaxy sample (SDSS-MGS) \citep{Ross2015}, the Baryon Oscillation Spectroscopic Survey data release 12 (BOSS-DR12) \citep{Alam2017} and from the WiggleZ Dark Energy Survey \citep{Kazin2014}. $\mathcal{D}_{V}(z)$, $\sigma_{\mathcal{D}}$ and $r_{s}^{fid}$ are given in Mpc. The $\sigma_{\mathcal{D}}$ are not shown here since the three WiggleZ measurements are correlated and we take the full covariance matrix in the analysis.}
         \label{bao_data}
\end{table}
\begin{figure}
   \centering
   \includegraphics[width=0.45\textwidth]{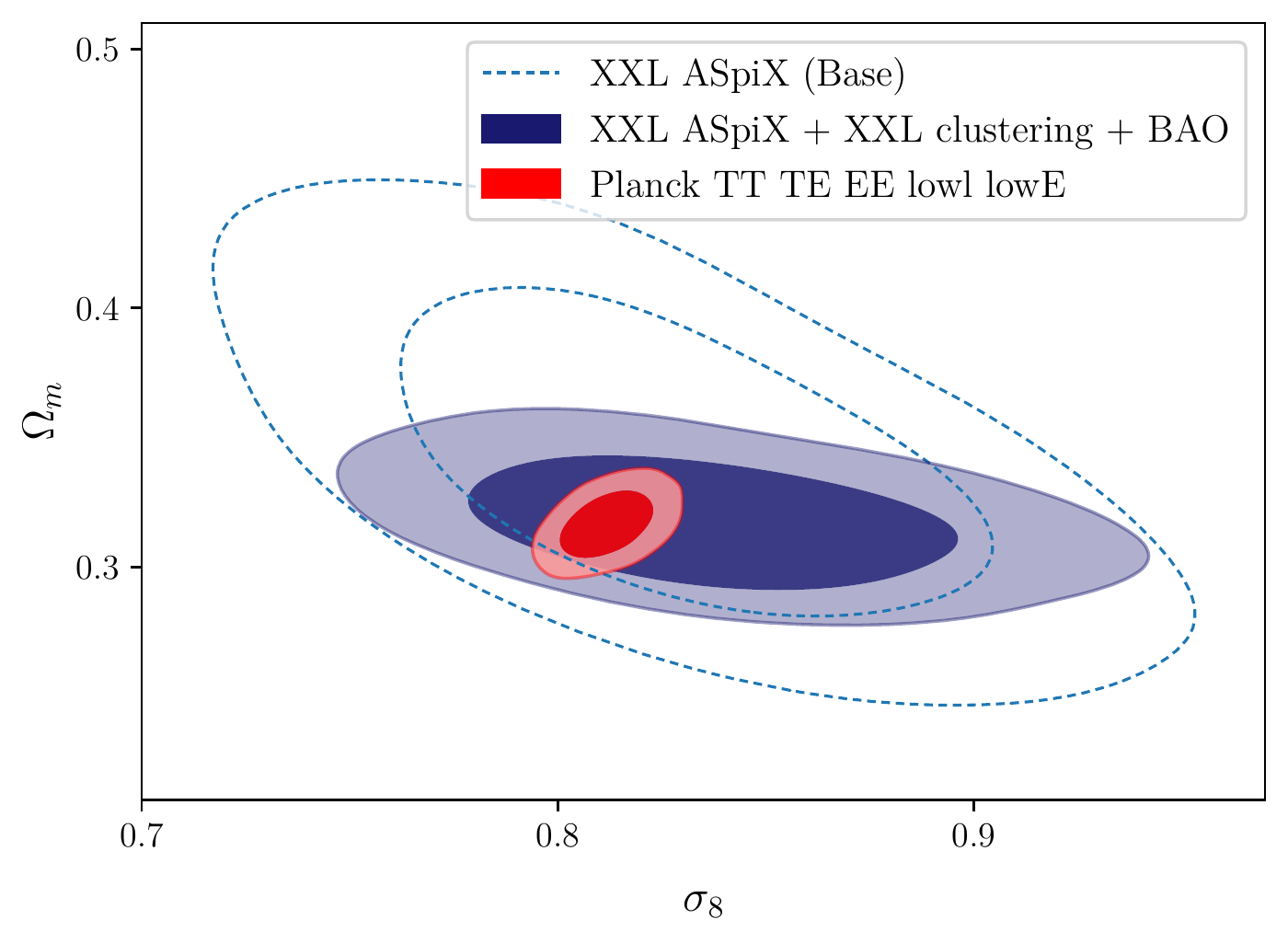} \hfill
   \caption{$\Omega_m-\sigma_8$ constraints obtained by  XXL ASpiX  alone (base) and XXL ASpiX + XXL clustering + BAO analysis and Planck-2018 (Planck TT TE EE lowl lowE).}
              \label{bao_aspix_joint}%
\end{figure}
In this section, we combine the ASpiX constraints  with those obtained from baryon acoustic oscillation (BAO) measurements of clustering of galaxies. The BAO data used in this analysis are reported in Table~\ref{bao_data}. We describe the adopted methodology  and  present the results from the joint analysis.  \\
We use two quantities to model the BAO distance measurements, (i) the spherically-averaged distance :
\begin{equation}
    D_{V}(z) = \left[ (1+z)^2 d_{a}^2(z) \frac{cz}{H(z)} \right]^{\frac{1}{3}}
\end{equation}
and (ii) the sound horizon at the drag epoch \citep{Eisenstein1998}:
\begin{equation}
    r_s(z_d) = \frac{2}{3 k_{eq}}\sqrt{\frac{6}{R(z_{eq})}}\ ln \left[ \frac{\sqrt{1+R(z_d)} + \sqrt{R(z_d) + R(z_{eq})}}{1 + \sqrt{R(z_{eq})}} \right]
\end{equation}
where $z_d$ is the redshift at the drag epoch, $z_{eq}$ is the matter-radiation equality redshift, $k_{eq}$ is the scale of the particle horizon at the equality epoch and $R(z)$ is the ratio of the baryon to photon momentum density at redshift $z$. We model these four quantities following \cite{Eisenstein1998} (with a CMB temperature of 2.725 K). The BAO  distance  is then given by :
\begin{equation}
    \mathcal{D}_{V}(z) = \frac{D_{V}(z)}{r_s(z_{d})}r_s^{fid}
\end{equation}
where $r_s^{fid}$ is the sound horizon computed for a chosen fiducial cosmology (see Table \ref{bao_data}). The likelihood is therefore modified by adding a Gaussian log-likelihood term in equation \ref{eq:lnLzi-2_sec} : 
\begin{equation}\label{eq:like_bao_term}
\begin{aligned}
    {\cal L}_{BAO} =& \frac{1}{2} \sum_i \left[ \frac{\mathcal{D}^{th}(z_i) - \mathcal{D}^{set\ I}_i}{\sigma_{\mathcal{D}^{set\ I}_i}} \right]^2  + \\
    &\frac{1}{2} \left[ \vec{\mathcal{D}}^{th} - \vec{\mathcal{D}}^{set\ II}\right] C^{-1}_{WiggleZ} \left[ \vec{\mathcal{D}}^{th} - \vec{\mathcal{D}}^{set\ II}\right] ,
\end{aligned}
\end{equation}
where the inverse covariance matrix, $C^{-1}_{WiggleZ}$, comes from the fact that the three WiggleZ measurements are correlated \citep[see Table 4 of][]{Kazin2014}.\\
\newline
Combining the BAO and the XXL 2PCF decreases the uncertainties by 62\% (FoM increased by a factor of 2.6) and confirms the agreement with Planck ($\Omega_m$ = $0.317\pm0.017$,  $\sigma_8$ = $0.845^{+0.035}_{-0.042}$ and $S_8=0.861^{+0.033}_{-0.042}$). The results are shown in Table~\ref{best_fit_cosmo_wfix}. The $\Omega_m-\sigma_8$ contours are shown in Fig.~\ref{bao_aspix_joint}. 

\subsection{XXL ASpiX + Planck CMB}
\begin{figure}
   \centering
   \includegraphics[width=0.45\textwidth]{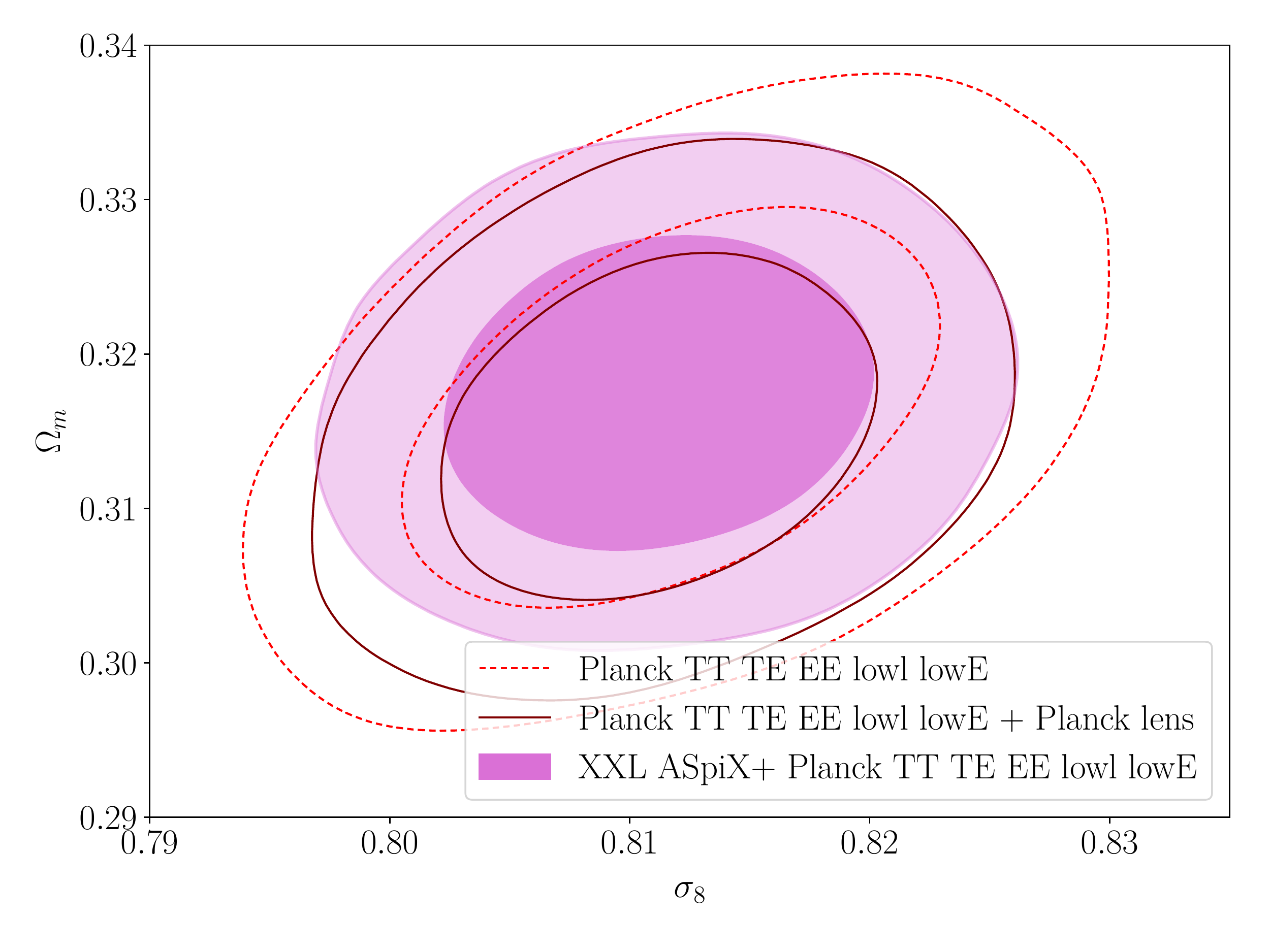} \hfill
   \caption{Comparison between Planck CMB + XXL ASpiX and Planck CMB + Planck lens and Planck CMB alone comparison of the $\Omega_m-\sigma_8$ constraints.}
              \label{planck_aspix_joint}%
\end{figure}
In this section, we  combine our results with Planck-2018 (CMB anisotropy measurement) by means of importance sampling. \\
The results are shown in Table~\ref{best_fit_cosmo_wfix}. The $\Omega_m-\sigma_8$ contours are shown in Fig.~\ref{planck_aspix_joint}.\\
Combining our results with Planck-2018 (CMB anisotropies measurement) we reduce Planck uncertainties by 30\% (FoM increased by a factor of 1.4) on $\Omega_m$ and $\sigma_8$ and find a good agreement with the constraints from the combination of Planck-2018 and Planck lensing (lensing potential analysis of the temperature and polarisation data), see Fig.~\ref{bao_aspix_joint}. We note that the XXL + Planck-2018 combination yields comparable constraints to the Planck-2018 + Planck lensing combination.

\section{Cosmological modelling with actualised scaling relations}\label{sec:mauro_laws}

\begin{table*}[]
\captionsetup{singlelinecheck = false, justification=justified}
\caption{Cluster scaling laws used in section \ref{sec:mauro_laws}.}
    \centering
    \begin{tabular}{lcccc}
         Law & $X_{0}$ & $\alpha$ & $\gamma$ & Scatter \\
          & & & & (natural log)\\
         \noalign{\smallskip}
         \hline
         \noalign{\smallskip}
         \hline
         \noalign{\smallskip}
         $T_{300kpc}-M_{500,WL}$ & $2.46 \pm 0.65$ keV & $0.85 \pm 0.39$ & $0.32 \pm 0.75$ & 0.13 \\
         \noalign{\smallskip}
         $L_{500,WL}^{XXL}-T_{300kpc}$ & ($20.9 \pm 5.0$) $\times 10^{41} \mathrm{erg.s^{-1}}$  & $2.63 \pm 0.34$ & $2.17 \pm 0.94$ & 0.38 \\
         \noalign{\smallskip}
         $\mathrm{r_{c}}-\mathrm{r_{500}}$ & 0.15 & - & - & 0.1 \\
    \end{tabular}
    \tablefoot{We disperse over the luminosity, temperature and core radius distributions in this case. Uncertainties on parameters indicate that these parameters are left free during the analysis.  The values shown in this table are calculated assuming $\Omega_m=0.28$ and $h=0.7$. During the analysis, the parameter means and covariances are rescaled as a function of cosmology as described in section \ref{sec:mauro_laws}.} 
    \label{scal_mauro_table}
\end{table*}
\begin{figure}
   \centering
   \includegraphics[width=0.45\textwidth]{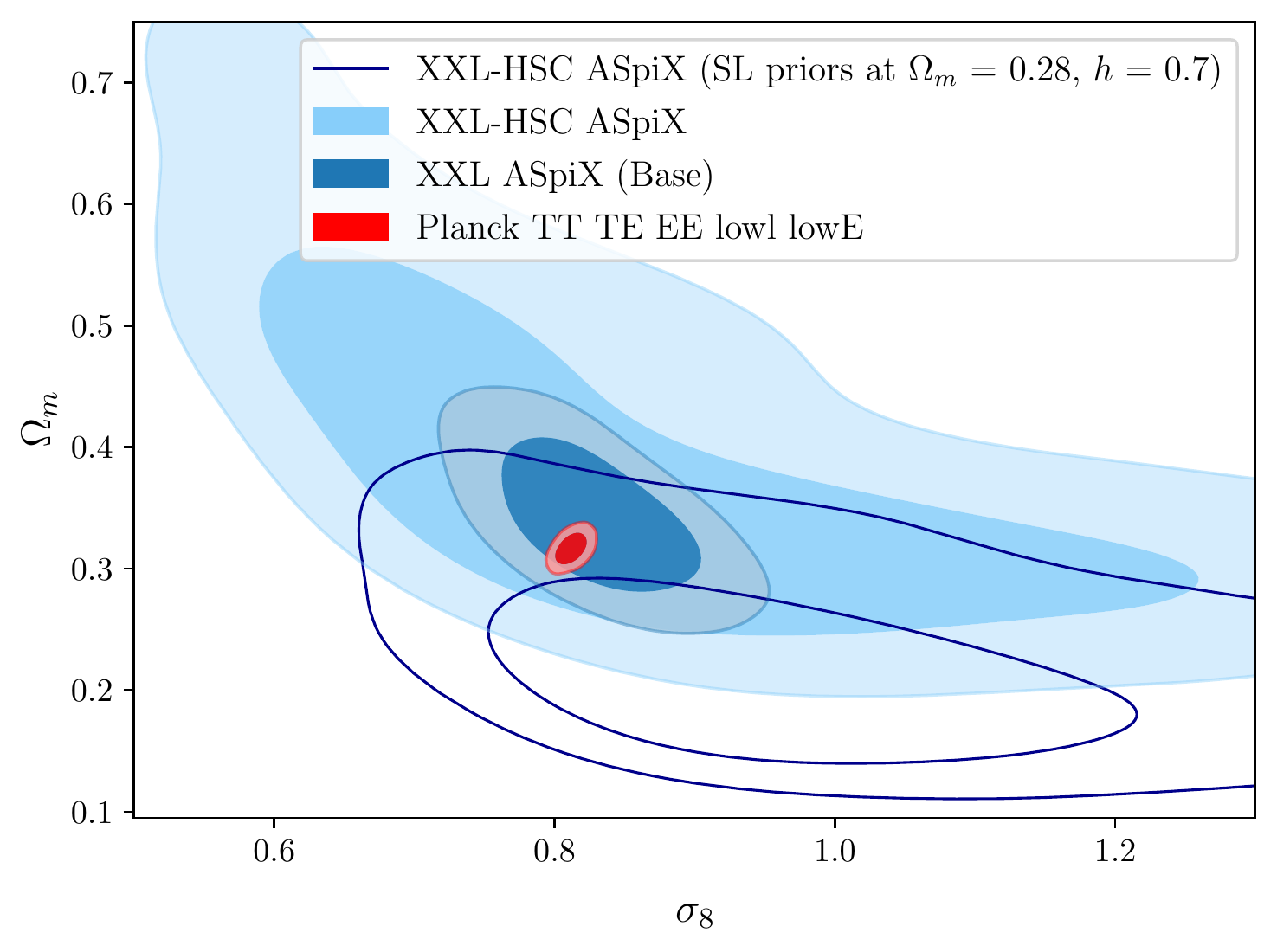} \hfill
   \caption{Impact of thawing parameters in the scaling relations. XXL ASpiX (base) refers to the results presented in section \ref{main_results} (2 free scaling parameters); XXL-HSC ASpiX contours are the results following the methodology presented in section \ref{sec:mauro_laws} (6 free, cosmology-dependent, scaling parameters); simple dark blue contours, same as XXL-HSC but the prior of the scaling coefficients are fixed to the indicated cosmology. We can see that in this case, the results in $\Omega_m$-$\sigma_8$ are shifted (lower $\Omega_m$ and higher $\sigma_8$) due to the fact that the scaling relation priors do not depend on the cosmology and then introduce a bias in the analysis.}
              \label{all_free_mauro_s8_om}%
\end{figure}
Two independent mass-observable studies (\citealp{Eckert2016}, hereafter XXL paper XIII, and \citealp{Umetsu2020}) suggest that cluster masses were overestimated in our first analysis based on the CFHT lensing data (XXL paper IV). In this section, we rerun the cosmological analysis, assuming our newly determined scaling relations from the joint XXL-HSC (Hyper Suprime-Cam Survey) analysis by \cite{Umetsu2020} and \cite{Sereno2020}.\\ 
\newline
For this purpose, we follow the formalism of \cite{Umetsu2020} for the scaling relations.\\
We consider the cluster true mass $\rm{M_{500,True}}$ as the fundamental property of galaxy clusters for the $T-M$ relation and we use the weak lensing mass $\rm{M_{500,WL}}$ as a mass proxy.\\
\cite{Umetsu2020} characterised the weak lensing mass bias as a
function of true cluster mass using cosmological $N$-body simulations
\citep[the dark-matter-only run from the BAHAMAS project;][]{McCarthy2017, McCarthy2018}.
They estimated that, at the typical mass for the XXL sample ($M_{500} = 7 \times 10^{13} h^{-1}M_{\odot}$), the bias is approximately -11\%. We therefore apply a correction for the weak lensing mass bias by assuming a constant of -11\%:
\begin{equation}
\frac{T_{300kpc}}{X_{0,T-M}} = \left(\frac{M_{500,True}}{7 \times
10^{13} h^{-1}M_{\odot}}\right)^{\alpha_{T-M}} \left(\frac{E(z)}{E(z=0.3)}\right)^{\gamma_{T-M}}
\end{equation}
with
\begin{equation}\label{eq:log_sl_keiichi}
\begin{aligned}
\mathrm{log_{10}}(M_{500,WL}) = &\mathrm{log_{10}}(M_{500,True}) \\
&+\mathrm{log_{10}} (1+b_{WL}) \pm \sigma_{\log_{10}M_\mathrm{WL}}
\end{aligned}
\end{equation} 
and we assume a Gaussian prior on $\mathrm{log_{10}} (1+b_{WL})$ of $\mathrm{log_{10}} (1+b_{WL}) =  \mathrm{log_{10}} (1-0.11) \pm 5\% / \mathrm{ln} 10$ to marginalise over the mass calibration uncertainty of $\pm 5\%$, see \cite{Umetsu2020}. Here $\sigma_{\log_{10}M_\mathrm{WL}}$ in Equation~\ref{eq:log_sl_keiichi} is the intrinsic scatter of weak-lensing mass at fixed true cluster mass, $M_\mathrm{500,True}$.\\
The $L-T$ relation is given by:
\begin{equation}
\frac{L^{XXL}_{500,WL}}{X_{0,L-T}} = \left(\frac{T_{300kpc}}{1~{\rm keV}}\right)^{\alpha_{L-T}} \left(\frac{E(z)}{E(z=0.3)}\right)^{\gamma_{L-T}}
\end{equation}
and we keep equation \ref{eq_scal_rc} for the relation between $r_c$ and $r_{500}$.\\
We fit the coefficients of the $T-M$ and $L-T$ relations (namely : \{$X_{0,T-M}$, $X_{0,L-T}$, $\alpha_{T-M}$, $\alpha_{L-T}$, $\gamma_{T-M}$, $\gamma_{L-T}$, $\sigma_{T-M}$, $\sigma_{\log_{10}M_\mathrm{WL}}$, $\sigma_{L-T}$\}, with $\sigma_{T-M/L-T}$ the log-normal intrinsic scatters) using the publicly available {\sc LIRA} package \citep{Sereno2016a,Sereno2016b} for the XXL C1 sample (using the procedure and measurements described in \citealp{Sereno2020} and \citealp{Umetsu2020}). The results, computed for $\Omega_m = 0.28$ and $h = 0.7$ in a flat $\Lambda$CDM universe, are shown in Table \ref{scal_mauro_table}.\\
\newline
The effective impact of the cosmological dependence\footnote{on $\Omega_m$ and $h$, for a flat $\Lambda$CDM model} of
 weak lensing mass measurements and luminosities is expected to be small given the parameter range considered and the statistical/systematics errors inherent to our cluster sample. Nevertheless, to ensure better consistency, we model - a posteriori - the effect of cosmology on the scaling relations as follows:\begin{itemize}
\item[•] We use an analytical approximation \citep{Sereno2015} to account for the dependence of the lensing mass on cosmology:
\begin{equation}
    M_{500,WL} \propto D_l^{-\frac{3\delta_{\gamma}}{2-\delta_{\gamma}}} \left(\frac{D_{ls}}{D_{s}}\right)^{-\frac{3}{2-\delta_{\gamma}}} H(z)^{-\frac{1+\delta_{\gamma}}{1-\delta_{\gamma}/2}}
\end{equation}
where $D_l$, $D_{s}$, $D_{ls}$ are the lens, the source and the lens-source angular diameter distances respectively. In a first approximation, we assume a linear relation between the cluster redshift and mean redshift of the source galaxies : 
\begin{equation} \label{equ_redshift_lin}
    \langle z_\mathrm{sources}\rangle = 0.714\ z_{cluster} + 0.786
\end{equation}
This results in a mean source-galaxy redshift of 1 for a cluster at 0.3 and of 1.5 for a cluster at redshift 1. We use $\delta_{\gamma} = 0.196$, fitted on our C1 sample, to rescale the masses on a grid of $\Omega_m$, and $h$ values. The ranges are defined to be $\Omega_m \in [0.1-0.8]$ and $h \in [0.5-0.9]$, with a 0.1 step for each parameter. This provides us with the masses for 40 combinations of $\Omega_m-h$ values. We then compute the T-M scaling relation for each [$\Omega_m-h$] points of the grid to obtain the corresponding mean values and covariance matrices.
\item[•]  Masses from the M-T relations are used to rescale   $r_{500}$ ($r_{500,rescale}$); we then extrapolate the luminosities within $r_{500,rescale}$ assuming a $\beta$-profile with a core radius $r_c = 0.15\ r_{500,rescale}$ and a slope $\beta = 2/3$.\\
Finally, luminosities are normalised  by the correction factor  $(d_L/d_L^{fid})^2$. This procedure provides us with rescaled luminosities for the 40 combinations of $\Omega_m-h$ values. We then compute the L-T scaling relation for each $\Omega_m-h$ point of the grid, to obtain corresponding mean values and covariance matrices. 
\item[•] Here, the cosmological analysis  deals with five free cosmological parameters : \{$\Omega_m,\ \sigma_8,\ \Omega_b,\ n_s,\ h$\} plus 6 free scaling relation parameters : \{$X_{0,T-M},\ X_{0,L-T},\ \alpha_{T-M},\ \alpha_{L-T},\ \gamma_{T-M},\ \gamma_{L-T}$\}. In the MCMC, the values of the six scaling relation parameters are limited through adaptive Gaussian priors, by interpolating the means and covariance matrices over the grid of 40 combinations of $\Omega_m-h$ values.
\item[•] We  disperse  temperatures and luminosities; these scatters are assumed to be independent of cosmology. We moreover introduce a log-normal scatter in the $r_c-r_{500}$ relation. 
\end{itemize}
Resulting constraints on $\Omega_m-\sigma_8$  (we will refer to this model as XXL-HSC ASpiX)  are shown in  Fig.~\ref{all_free_mauro_s8_om} and compared with the results of Sec. \ref{main_results}  (referring to this model as the base model). The constraints when adding the XXL 2PCF and BAO measurements are shown in Fig. \ref{all_free_mauro_s8_om_combined}. All the results are shown in Table~\ref{best_fit_cosmo_wfix_sl_adapt}.\\
We find, for XXL-HSC ASpiX, $\Omega_m$ and $\sigma_8$ results slightly higher and with larger error bars, $\Omega_m$ = $0.378^{0.068}_{0.13}$, $\sigma_8$ = $0.89^{0.12}_{0.28}$ ($S_8 = 0.970^{+0.067}_{-0.21}$). Nevertheless, the results are compatible at the 1-$\sigma$ level with our base model and the Planck CMB. Of course, since our uncertainties are now larger, we are compatible with the Planck S-Z cluster counts as well.\\ 
Adding the XXL clustering, we find a smaller $\Omega_m$ = 0.296 $\pm$ 0.034 and a higher $\sigma_8$ = $0.99^{+0.14}_{-0.23}$ ($S_8 = 0.98^{+0.11}_{-0.21}$) letting us fully consistent with Planck CMB at the 1-sigma level.\\
Combining XXL-HSC ASpiX with the XXL clustering and the BAO measurements, the results are shifted ($\Omega_m$ = 0.364 $\pm$ 0.015, $\sigma_8$ = $0.793^{0.063}_{0.12}$, $S_8 = 0.872^{+0.068}_{-0.12}$) and we find results in a better agreement with the Planck S-Z cluster sample while being consistent with Planck CMB at 2.2$\sigma$.

\begin{figure}
   \centering
   \includegraphics[width=0.45\textwidth]{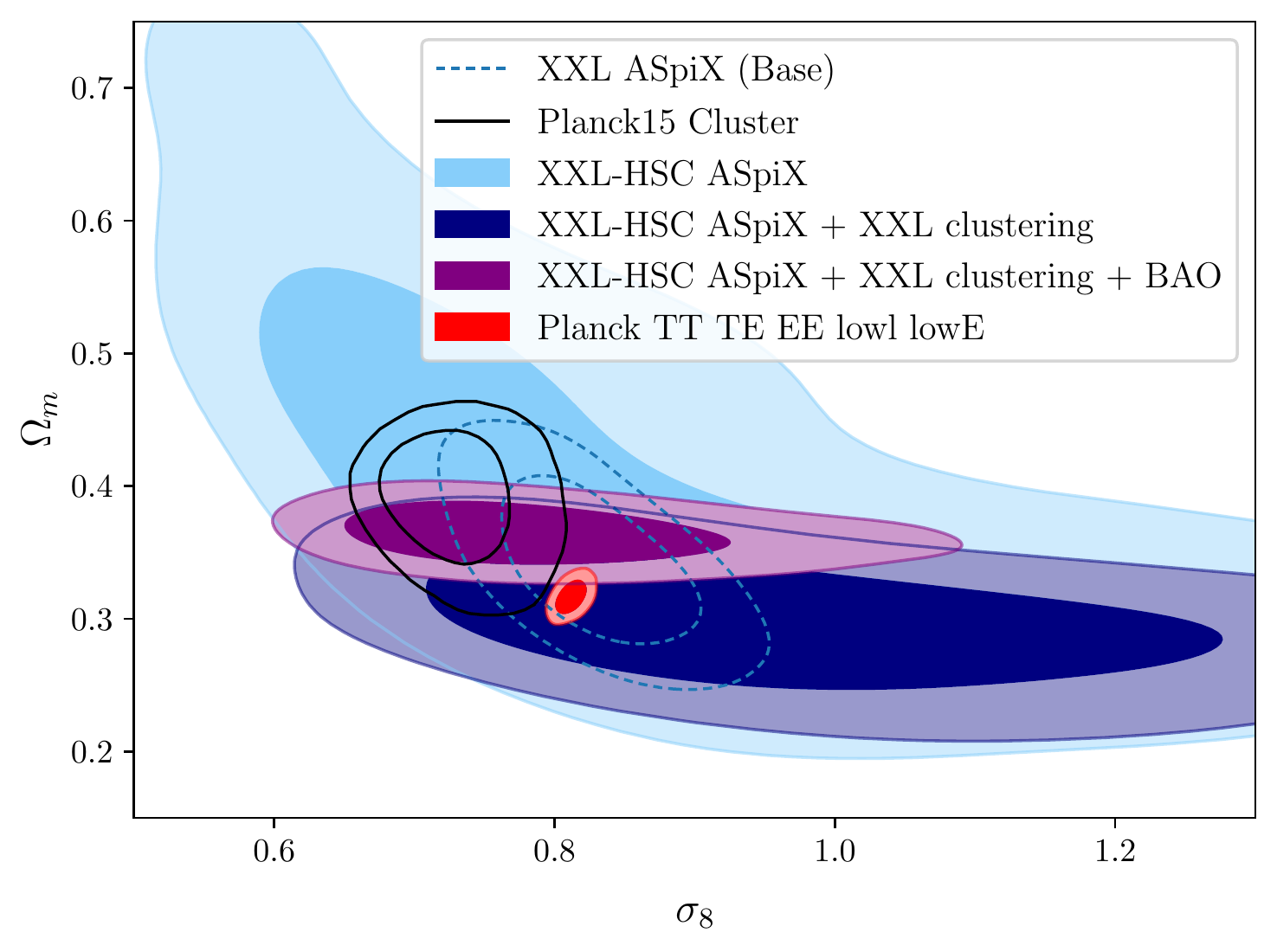} \hfill
   \caption{ 
Same as Fig. \ref{all_free_mauro_s8_om} when adding constraints from XXL clustering and  external BAO. The dashed line stands for the case where only two scaling coefficients are let free. 
   }
              \label{all_free_mauro_s8_om_combined}%
\end{figure}
\begin{table*}
\captionsetup{singlelinecheck = false, justification=justified}
    \caption{ASpiX cosmological constraints for the HSC-XXL ASpiX model and the joint analysis (flat $\Lambda$CDM).}
     $$ 
         \begin{tabular}{*{5}{P{2.3cm}}}
            \hline
            \noalign{\smallskip}
            \tiny Parameter & \tiny XXL-HSC ASpiX & \tiny XXL-HSC ASpiX & \tiny XXL-HSC ASpiX & \tiny Priors\\
             & & \tiny + XXL clustering & \tiny + XXL clustering &\\
              &  &   & \tiny + BAO &\\
            \noalign{\smallskip}
            \hline
            \noalign{\smallskip}
            $\Omega_m$\dotfill&\small $0.378^{+0.068}_{-0.130}$ &\small $0.296 \pm 0.034$ &\small $0.364 \pm 0.015$ & \small $\mathcal{U}(0.09,1.0)$\\
            \noalign{\smallskip}
            $\sigma_8$\dotfill &\small $0.890^{+0.120}_{-0.280}$ &\small $0.990^{+0.140}_{-0.230}$ &\small $0.793^{+0.063}_{-0.120}$ & \small $\mathcal{U}(0.05,2.0)$\\
            \noalign{\smallskip}
            $S_8$\dotfill &\small $0.970^{+0.067}_{-0.210}$ &\small $0.980^{+0.110}_{-0.210}$ &\small $0.872^{+0.068}_{-0.120}$ & -\\
            \noalign{\smallskip}
            $\Omega_b$\dotfill &\small $0.049 \pm 0.002$ &\small $0.047 \pm 0.001$ &\small $0.047 \pm 0.001$ & \small $\mathcal{N}(0.0493,0.0035^2)$\\
            \noalign{\smallskip}
            $h$\dotfill &\small $0.674^{+0.017}_{-0.019}$  &\small $0.693 \pm 0.010$ &\small $0.682 \pm 0.009$ & \small $\mathcal{U}(0.55,0.9)$\\
            \noalign{\smallskip}
            $n_s$\dotfill &\small $0.965 \pm 0.015$ &\small $0.964 \pm 0.008$ &\small $0.964 \pm 0.008$ & \small $\mathcal{N}(0.9649,0.022^2)$\\
            \noalign{\smallskip}
            \hline
         \end{tabular}
     $$ 
      \tablefoot{$S_8$ is defined to be  $S_8=\sigma_8(\Omega_m / 0.3)^{0.5}$. We do not quote constraints on the nuisance parameters used in this analysis. Furthermore, the mean and covariance of the Gaussian priors for the 6 free scaling relation parameters (namely \{$X_{0,T-M},\ X_{0,L-T},\ \alpha_{T-M},\ \alpha_{L-T},\ \gamma_{T-M},\ \gamma_{L-T}$\}) are not shown here since they are rescaled as a function of cosmology (described in Section \ref{sec:mauro_laws}).}
         \label{best_fit_cosmo_wfix_sl_adapt}
\end{table*}

\section{Discussion}\label{discussion_section}

\subsection{Result summary} 

\begin{figure}
   \centering
   \includegraphics[width=0.45\textwidth]{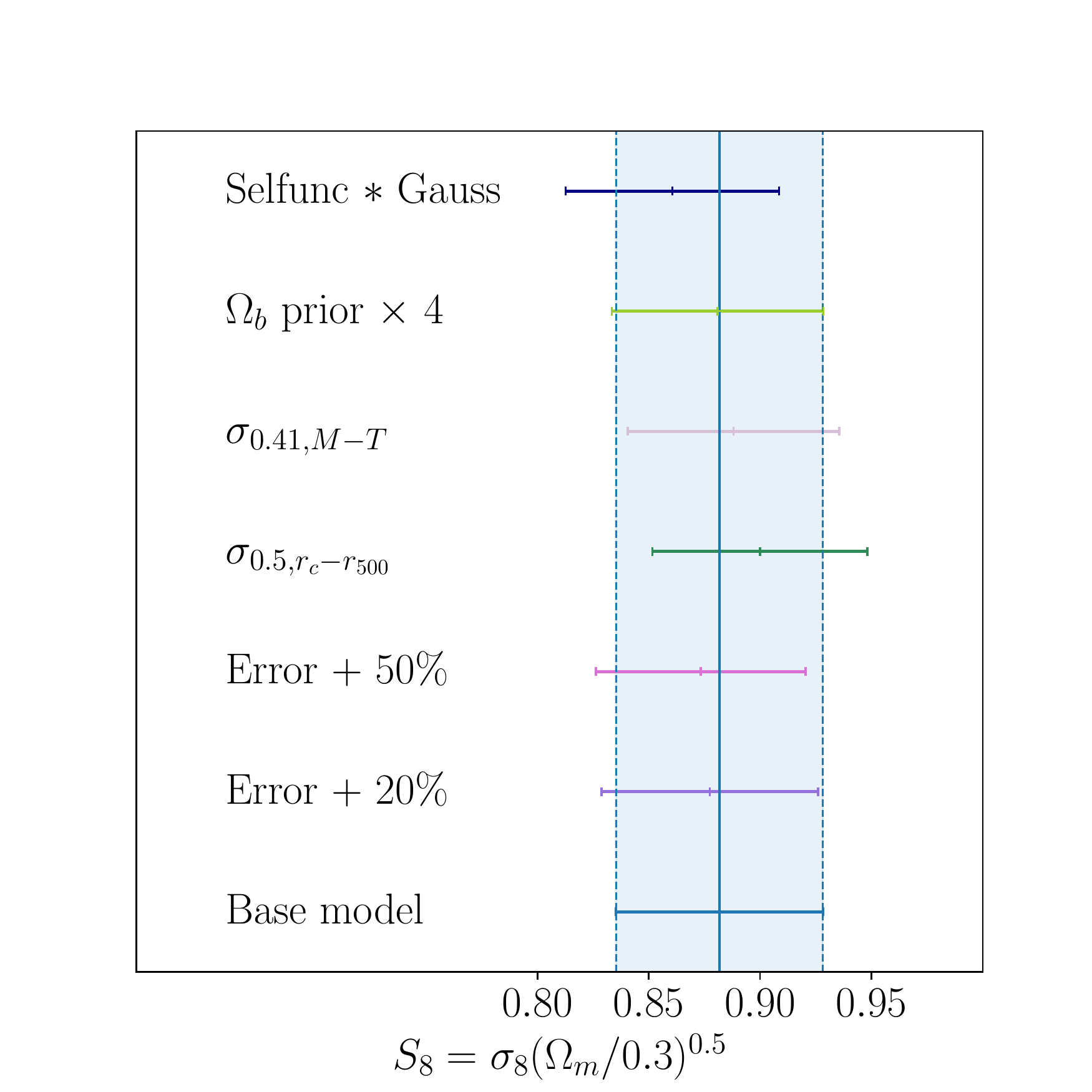} \hfill
   \caption{$S_8= \sigma_8(\Omega_m / 0.3)^{0.5}$ results for all tests discussed in section \ref{discussion_section}}
              \label{discussion_all}
\end{figure}

Fig. \ref{fixw_sep} shows that, as expected, the constraints on $\Omega_m-\sigma_8$ have improved by more than a factor of two with respect to XXL paper XXV, under exactly the same hypotheses. This confirms on real data, the power of the ASpiX forward modelling in terms of simplicity and accuracy. The size of the error bars is now comparable to that from the Planck S-Z cluster sample \citep{Planck2016}. At this point, it is important to recall that the Planck cluster sample contains almost three times as many clusters as XXL, that these clusters are much more massive and that the XXL scaling relations do not rely on external cluster calibration samples, or on hydrodynamical simulations, contrary to Planck's; currently, the two data set are still consistent at the 2-$\sigma$ level, even though the favoured XXL cosmology is closer to the Planck CMB values. Combining the ASpiX constraints with the results from the cluster-cluster correlation function from the same sample improves the constraints by  23\% (Fig. \ref{clustering_aspix_joint}), and by 62 \% (Fig. \ref{bao_aspix_joint}) when adding both cluster-cluster correlation function and BAO measurements.\\
In a second step, we have rerun the ASpiX analysis by implementing  the actualised scaling relations along with a modelling of their cosmological dependence, i.e. by increasing the degrees of freedom on cluster physics, from  2 to 6 (Fig. \ref{all_free_mauro_s8_om} and Table \ref{scal_mauro_table}). As expected, this results in larger error bars: we now favour a higher $\sigma_8$ value, $\sigma_8 = 0.89^{+0.12}_{-0.28}$, but we are still compatible at the 1-$\sigma$ with the Planck CMB.\\
\newline
At this stage, we are close to having exhausted the cosmological information contained in the current data-set relative to the XXL C1 cluster sample. It is instructive to review and discuss the various possible sources of systematic error that impinge upon these new results and, thus, assess their robustness.\\
A number of possible systematic uncertainties were already identified and discussed in the previous cosmological analysis of XXL paper XXV. In the following section, we review further hypotheses and examine the impact on the initial base model. To quantify the robustness of our results, we analyse the posterior distribution of the $S_8= \sigma_8(\Omega_m / 0.3)^{0.5}$ product.

\begin{table*}
\captionsetup{singlelinecheck = false, justification=justified}
\caption{Posterior agreement between the various cases studied in Sec. \ref{discussion_section}.}
\centering
     $$ 
         \begin{tabular}{|l|*{7}{P{1.75cm}|}}
            \hline
            
            \tiny Models  &\tiny Base &\tiny Error + $20\%$ & \tiny Error + $50\%$ &\tiny $\sigma_{0.5,r_c-r_{500}}$ & \tiny $\sigma_{0.41,M-T}$ & \tiny $\Omega_b$ prior $\times$ 4 & \tiny $\mathrm{Selfunc} \ast \mathrm{Gauss}$\\
            \hline
            \noalign{\smallskip}
            \hline
            \tiny Base  & \cellcolor{gray}& \cellcolor{blue!10} \tiny $0.03\sigma$ & \cellcolor{blue!10}\tiny $0.03\sigma$ & \cellcolor{blue!10} \tiny $0.06\sigma$ & \cellcolor{blue!10} \tiny $0.1\sigma$& \cellcolor{blue!10} \tiny $0.02\sigma$ & \cellcolor{blue!10} \tiny $0.2\sigma$\\
            \hline
            \tiny Error + $20\%$ & \tiny $0.04\sigma$ & \cellcolor{gray} &\cellcolor{blue!10} \tiny $0.03\sigma$ &\cellcolor{blue!10} \tiny $0.04\sigma$  & \cellcolor{blue!10} \tiny $0.1\sigma$ & \cellcolor{blue!10} \tiny $0.03\sigma$ & \cellcolor{blue!10} \tiny $0.2\sigma$\\
            \hline
            \tiny Error + $50\%$ &  \tiny $0.05\sigma$ & \tiny $0.03\sigma$  & \cellcolor{gray} &\cellcolor{blue!10} \tiny $0.04\sigma$ & \cellcolor{blue!10} \tiny $0.08\sigma$ & \cellcolor{blue!10} \tiny $0.02\sigma$ & \cellcolor{blue!10} \tiny $0.1\sigma$\\
            \hline
            \tiny $\sigma_{0.5,r_c-r_{500}}$ & \tiny $0.04\sigma$ & \tiny $0.1\sigma$  & \tiny $0.1\sigma$ & \cellcolor{gray} & \cellcolor{blue!10} \tiny $0.1\sigma$ & \cellcolor{blue!10} \tiny $0.03\sigma$ & \cellcolor{blue!10} \tiny $0.3\sigma$\\
            \hline
            \tiny $\sigma_{0.41,M-T}$ & \tiny $0.3\sigma$ & \tiny $0.3\sigma$  & \tiny $0.3\sigma$ & \tiny $0.4\sigma$ & \cellcolor{gray} & \cellcolor{blue!10} \tiny $0.09\sigma$ & \cellcolor{blue!10} \tiny $0.05\sigma$\\
            \hline
            \tiny $\Omega_b$ prior & \tiny $0.04\sigma$ & \tiny $0.03\sigma$ & \tiny $0.04\sigma$ & \tiny $0.09\sigma$ & \tiny $0.3\sigma$ & \cellcolor{gray} & \cellcolor{blue!10} \tiny $0.15\sigma$\\
            \hline
            \tiny $\mathrm{Selfunc} \ast \mathrm{Gauss}$ & \tiny $0.5\sigma$ & \tiny $0.5\sigma$ & \tiny $0.5\sigma$ & \tiny $0.7\sigma$ & \tiny $0.1\sigma$ & \tiny $0.4\sigma$ & \cellcolor{gray}\\
            \hline
            \hline
         \end{tabular}
     $$ 
      \tablefoot{Left of the diagonal: the posterior agreement in the $\Omega_{m}-\sigma_8$ plane. Right of the diagonal (blue cells): the posterior agreement in the 4D $\Omega_m-\sigma_8-\Omega_b-n_s$ plane.}
         \label{agreement_discuss}
\end{table*}
\subsection{Impact of error model} 

To study the impact of the error model, we arbitrarily modify the true error model by increasing errors by 20 and 50\%, and we monitor the effect on the cosmological constraints (referring to them as Error + $20\%$ and Error + $50\%$ respectively).\\
We can see from Fig.~\ref{discussion_all} that increasing the relative measurement errors slightly increases the uncertainty on $S_8$ without any drastic change in the mean result. Nevertheless, the general pattern seems to indicate that assuming excessive measurement errors tends to decrease $S_8$. The agreement between models is shown in Table~\ref{agreement_discuss}.

\subsection{Scaling relation model}\label{subsection_discuss_sr}

We now investigate how the results are impacted by different scaling relation models. First, we present the results when only adding a scatter of 0.5 in the $r_c-r_{500}$ relation. Then we study the effect of adding scatter in the $M-T$ relation.\\
\newline
From now on, we will refer to the 0.5 scatter in the $r_c-r_{500}$ relation as the $\sigma_{0.5,r_c-r_{500}}$ model. In Fig.~\ref{discussion_all}, we can see that adding a 50\% scatter in the $r_c-r_{500}$ relation favours a slightly higher $S_8$ value while increasing the uncertainties by only 3\% compared to the base model. All in all, the results appear little affected (Table~\ref{agreement_discuss}). If we had added $\theta_c$ as the 4th dimension in the XOD  and used it in the cosmological inference, the error bars would probably have been smaller, but more dependent on the scatter value \citep[cf][]{Valotti2018}. \\
\newline
In the base model, we did not implement a scatter in the $M-T$ relation to keep the same configuration as in XXL paper XXV (all scatter is supposed to be encapsulated in L-T). However, because $\mathrm{HR}$ directly depends on cluster temperature, it is logical to include a dispersion. We then include a 0.41 scatter in the $M-T$ relation obtained from XXL paper IV. We will refer to this model from now on as $\sigma_{0.41,M-T}$.\\
This model is in good agreement with the base one with a significance level of posterior agreement in the $\Omega_m-\sigma_8-\Omega_b-n_s$ space of $0.1\sigma$ and $0.3\sigma$ in the $\Omega_m-\sigma_8$ plane, see Table~\ref{agreement_discuss} and Fig.~\ref{discussion_all}. 

\subsection{Impact on $\Omega_b$’s priors}

To ensure that the prior chosen for $\Omega_b$ is not too restrictive, we apply a Gaussian prior centred on the Planck-2018 values but with errors multiplied by a factor of 20 : $\Omega_b = 0.0493 \pm 0.015$ (i.e. our previous prior multiplied by 4). We will refer to this model from now on as $\Omega_b\ \mathrm{prior} \times 4$.\\
The resulting constraint on $S_8$  is shown in Fig.~\ref{discussion_all} and the agreement between models in Table~\ref{agreement_discuss}.\\
We find that the results are fully consistent with the base model (0.02$\sigma$ posteriors agreement in 4D $\Omega_m-\sigma_8-\Omega_b-n_s$ space and 0.04$\sigma$ posteriors agreement in $\Omega_m-\sigma_8$ only, see Table~\ref{agreement_discuss}). To conclude, because $\Omega_b\ \mathrm{prior} \times 4$ becomes computationally expensive, we determine that it is relevant to keep the base model prior for $\Omega_b$. 

\subsection{Error on the selection function}

In this section, we study the impact of uncertainties on the selection function. This is a priori a key issue because an ill-determined selection directly biases the modelling of the cluster number counts. 
Currently, our selection is based on simulations assuming spherically symmetric and $\beta=2/3$ profiles for the cluster emission.
To quantify the impact of a poorly monitored cluster selection on the cosmological inference, we degrade the selection function; i.e. we blur the current function displayed in Fig.~\ref{selfunc}  by a 2D adaptive Gaussian filter characterised by:
\begin{equation}
    \sigma_{sel.\ func.} = 0.05\  \mathrm{CR}^{-0.6} \theta_c^{-0.4}
\end{equation} As easily understandable, in this way, fainter and smaller clusters are more affected. We will refer to this modelling from now on as $\mathrm{Selfunc} \ast \mathrm{Gauss}$.\\
The $\mathrm{Selfunc} \ast \mathrm{Gauss}$ result on $S_8$ is shown in Fig.~\ref{discussion_all} and the agreement between models is shown in Table~\ref{agreement_discuss}. While increasing  uncertainties on $S_8$, the blurred selection function also lowers the mean $S_8$ value.

\subsection{Remaining sources of uncertainty}

In addition to the sources of systematic uncertainties reviewed above, we also note the main assumptions used in the course of the present study. Firstly, the covariance between the observable parameters (CR, HR and $\theta_c$) is neglected; the model has been slightly extrapolated in order to account for objects scattered out or in the measured domain. Furthermore, we do not consider the covariance between the scatters of the M-T and L-T scaling relations. In both cases, the scatter is assumed to be independent of the underlying cosmology. In the lensing analysis, we assume a linear relation between lens cluster redshift and the galaxy source photometric redshifts as stated in Eq.~\ref{equ_redshift_lin}. Finally, we restrict our analysis to only one particular mass function (Sec. \ref{sec:numerical_inputs}) for this study. We aim to examine the impact of these assumptions in the subsequent and final XXL analysis - consisting of a larger number of clusters - to determine the most accurate, unbiased cosmological estimates from the XXL sample.

\section{Conclusions}
Following \cite{Clerc2012b} and simulation case studies, we present the first application of the ASpiX cosmological forward modelling on real data with redshift information. The outcome confirms the flexibility and efficiency of the method.
The constraints obtained from the 178 XXL C1  clusters, under various hypotheses, yield a precision comparable to that of the current BAO and Planck S-Z samples, as shown in Figure \ref{all_free_mauro_s8_om_combined}. Nevertheless, the number of degrees of freedom left in the analysis reflect the accuracy of the recovered cosmological parameters, e,g. by comparing the XXL base model alone (dashed blue contours) to the tightest constraints from this analysis (purple contours). \\
In short, the current results present an improvement by a factor of two compared to the preceding $dn/dz$ analysis of the same sample. At this stage, we may recall the final cosmological modelling of the REFLEX survey number counts. It is based on the luminosity function of more than 800 clusters detected in the ROSAT All-Sky Survey \citep[$z<0.4$;][]{Boehringer2014}. Our base model analysis (Sec.~\ref{main_results}) on the 178 C1 clusters includes four free cosmological parameters plus two scaling relation coefficients as nuisance parameters; the REFLEX analysis let only the slope of the M-L relation free and assumed that the luminosity function does not evolve. Under these conditions, we find a precision on $\Omega_m$ comparable to that of REFLEX but almost twice as better for $\sigma_8$; both parameter sets being compatible within the error bars.\\
Another cosmological analysis of RASS clusters has been conducted by the ‘Weighing the Giants’ project.  The 224 ‘Giants’ are massive clusters spanning the $0<z<0.3$ redshift range. Gas masses from deep ROSAT and Chandra observations could be subsequently derived for 94 of them. Independent mass calibration was achieved by weak gravitational lensing for 27. This enabled the derivation of uniquely well-defined scaling relations and, subsequently, yielded a precision of the order of 5\% on $\sigma_8$ and $\Omega_m$ \citep[with standard priors on $\Omega_b$, $h$, and $n_s$ fixed;][]{Mantz2015}. Constraints are tighter than with the XXL clusters, but it is important to recall here that the only X-ray information used in the current study is the XMM 10ks survey data, which means a median number of photons of $\sim 200$ per cluster.  We can anticipate that devoting very large amounts of X-ray follow-up time to the XXL clusters would outperform the WtG constraints, thanks to the wider redshift range spanned by the XXL clusters. \\
Ultimately, the XXL XMM observation set will be reprocessed at full depth i.e. by running the detection algorithm on the mosaicked data \citep[][hereafter XXL paper XXIV]{Faccioli2018}. This will not only increase the sensitivity but also the surveyed area, since the current cluster catalogue (XXL paper XX)  was extracted only from the single pointings, restricted to an off-axis distance of 13 arcminutes.   
It is thus expected that the final XXL cosmological release will involve a sample twice as large as the current one, with a deeper C1 and C2 population. In the subsequent cosmological analysis, we shall add information from the third X-ray observable, the apparent core radius \citep{Valotti2018}. The final cosmological sample should bring an improvement of a factor $\sim 1.5-2$ on the present constraints. \\
Using the same sample of 178 clusters, our immediate next study will focus on the $w$ parameter of the $\Lambda$CDM model. To this purpose, we shall make use of the HSC full depth information on the background galaxy photometric redshifts; the cosmological dependence of the cluster lensing masses will be rescaled as a function of $w$. The inclusion of the cluster 2-point correlation function, while having little effect on the current study limited to $\Omega_m$ and $\sigma_8$, is expected to reduce the uncertainty on $w$ by a factor of two \citep{Pierre2011}. Similarly, we shall allow for more flexibility in the determination of the cluster selection function: by considering a range of cluster ellipticities and quantifying the impact of cool cores or central AGN in the detection, we shall be in a position to assess more precisely systematic uncertainties. \\
Because photometric redshifts are almost as efficient as spectroscopic redshifts in ASpiX \citep{Clerc2012a}, the application of the method to the up-coming eROSITA cluster sample should readily reveal  most of the eROSITA sample's cosmological potential. 

\section*{Acknowledgements}

XXL is an international project based around an XMM Very Large Programme surveying two 25 ${\rm deg^2}$ extragalactic fields at a depth of ${\rm \sim 6\ \times\ 10^{-15}\ erg\ cm^{-2}\ s^{-1}}$ in the [0.5-2] keV band for point-like sources. The XXL website is \url{http://irfu.cea.fr/xxl}. \\
The Saclay team acknowledges long term support from the Centre National d'Etudes Spatiales.\\
This work was supported by the Programme National Cosmology et Galaxies (PNCG) of CNRS/INSU with INP and IN2P3, co-funded by CEA and CNES.\\
MS acknowledges financial contribution from contract ASI-INAF n.2017-14-H.0 and INAF `Call per interventi aggiuntivi a sostegno della ricerca di main stream di INAF'. \\
LM acknowledges the grants PRIN-MIUR 2017 WSCC32 and  ASI-INAF n. 2018-23-HH.0.\\
KU acknowledges support from the Ministry of Science and Technology of Taiwan (grants MOST 106-2628-M-001-003-MY3 and MOST 109-2112-M-001-018-MY3) and from the Academia Sinica Investigator Award (grant no. AS-IA-107-M01).\\
Finally, the authors would like to thanks the referee for the useful comments.

\bibliographystyle{aa}
\bibliography{bibliography}

\begin{appendix} 
\newcommand{\be}{\begin{equation}}
\newcommand{\ee}{\end{equation}}
\newcommand{\ba}{\begin{eqnarray}}
\newcommand{\ea}{\end{eqnarray}}
\newcommand{\Mpl}{M_{\textrm{Pl}}}

\section{X-ray parameter measurements}
\label{measurement_sec}

The use of the XOD for cosmology requires accurate measurements along with realistic error estimates. In this section, we provide details on the various steps of the X-ray analysis for single clusters. Measurements are performed on the mosaicked co-added observations.

\subsection{Step 1: Manually monitoring $\theta_{bkg}$}

We use the interactive {\tt fluxmes} procedure \citep{Clerc2012b} to determine the cluster-centric distance from which the background can be safely estimated ($\theta_{bkg}$). In addition, this procedure allows us to check that all neighbouring sources have been correctly flagged by the pipeline; and possibly to re-adjust the corresponding masks. The {\tt fluxmes} application is based on a curve of growth analysis and decides where to set the "no man's land limit" around the cluster.  In the following, the particle background is measured  using a very large number of "closed-filter" observations, scaled to our XXL observations using the detector parts not exposed to the sky (corners); this results in two 25 \dd\ maps, corresponding  to the mosaicked observations and exposure maps.  The astrophysical background (passing through the XMM optics) is locally estimated in the [$\theta_{bkg}$, 15 arcmin] annulus around each cluster in the course of the {\tt pyproffit} procedure (see Sec. \ref{proffit}).

\begin{figure*}
   \centering
   \subfloat[][]{\includegraphics[width=0.33\textwidth]{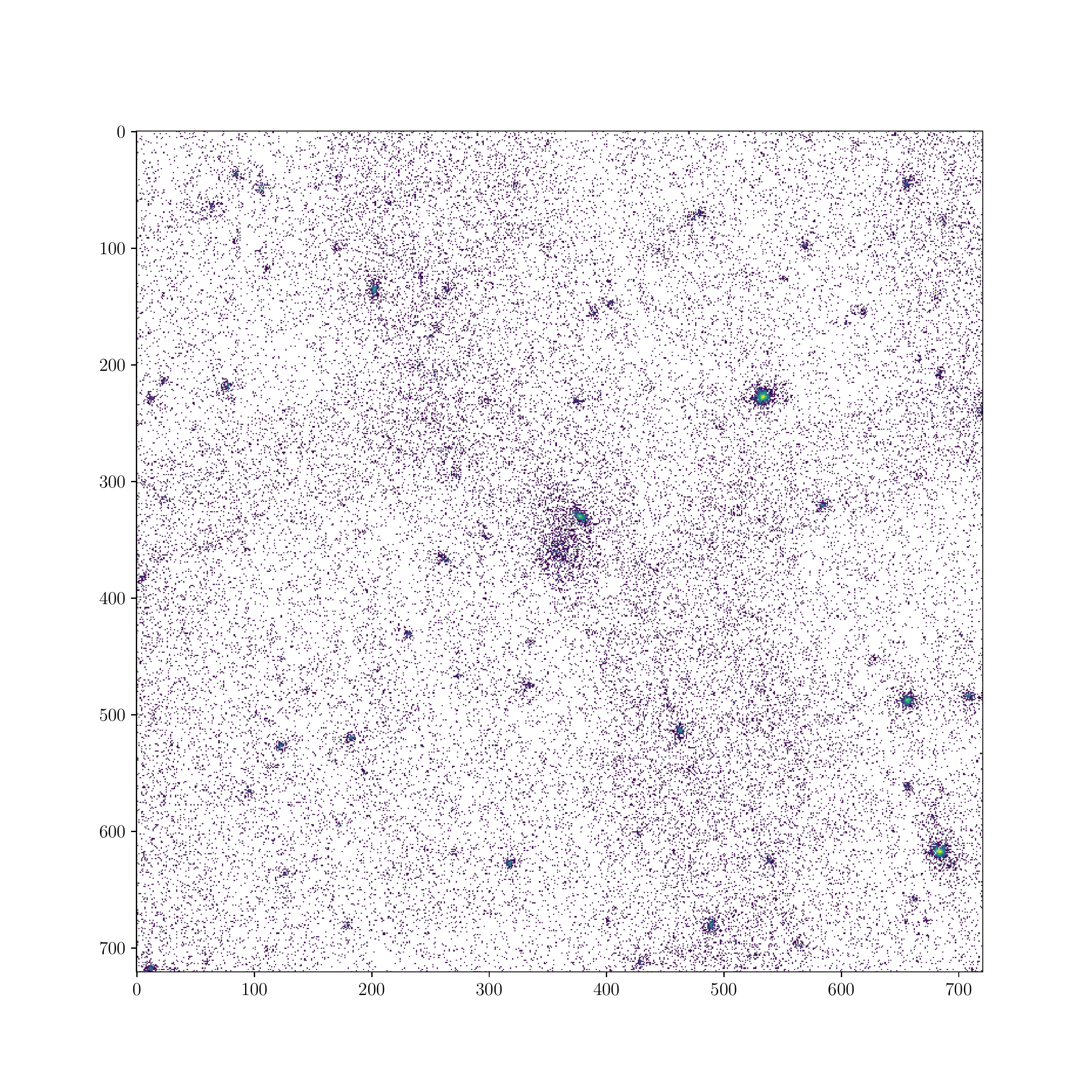}\label{mosaic_n0234}} \hfill
   \subfloat[][]{\includegraphics[width=0.33\textwidth]{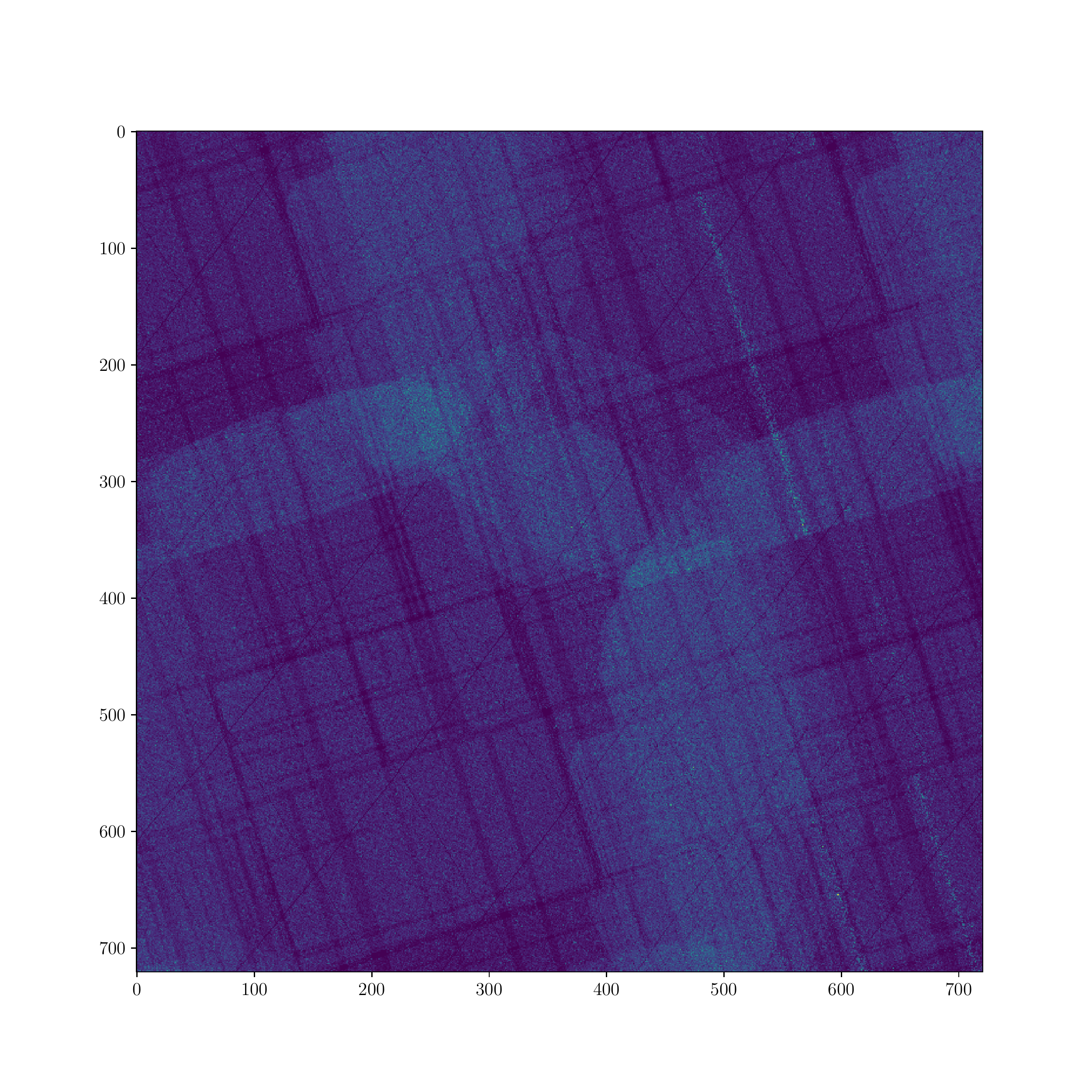}\label{expo_n0234}} \hfill
   \subfloat[][]{\includegraphics[width=0.33\textwidth]{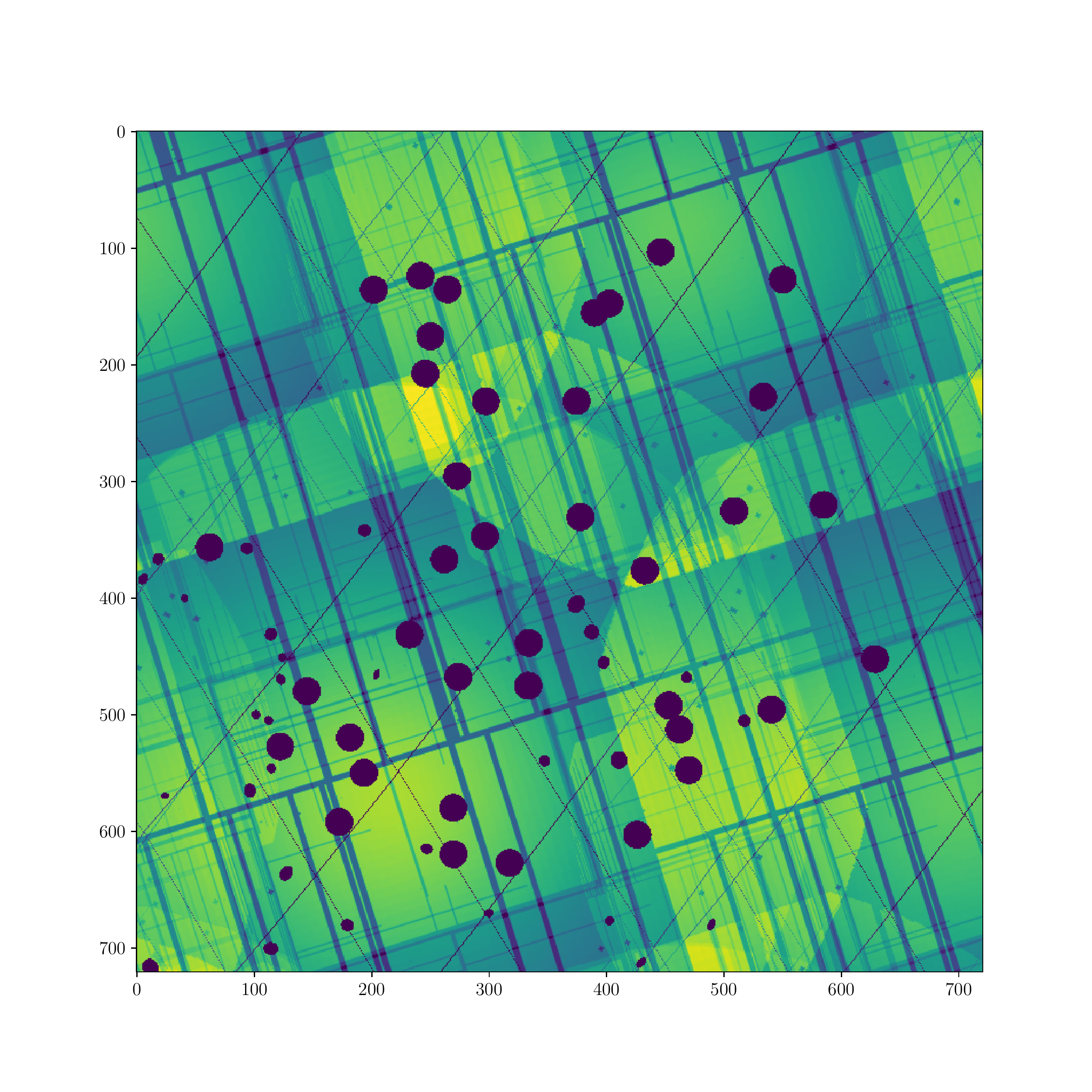}\label{mask_n0234}} \hfill\\
   \subfloat[][]{\includegraphics[width=0.33\textwidth,trim= 0cm 0cm 0cm .41cm]{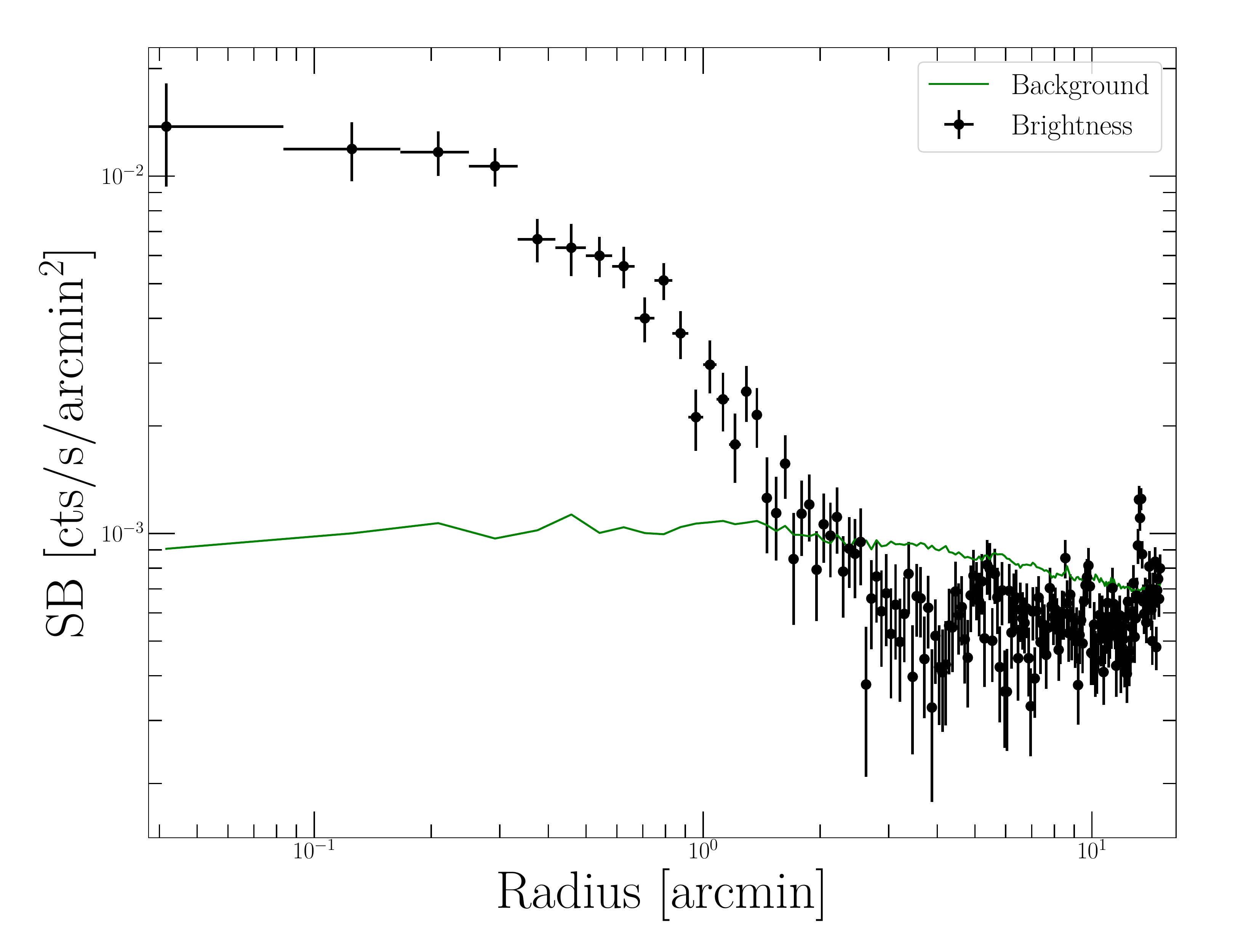}\label{profile_n0234}}\hfill
   \subfloat[][]{\includegraphics[width=0.33\textwidth,trim= 0cm 0.39cm 0cm 0cm]{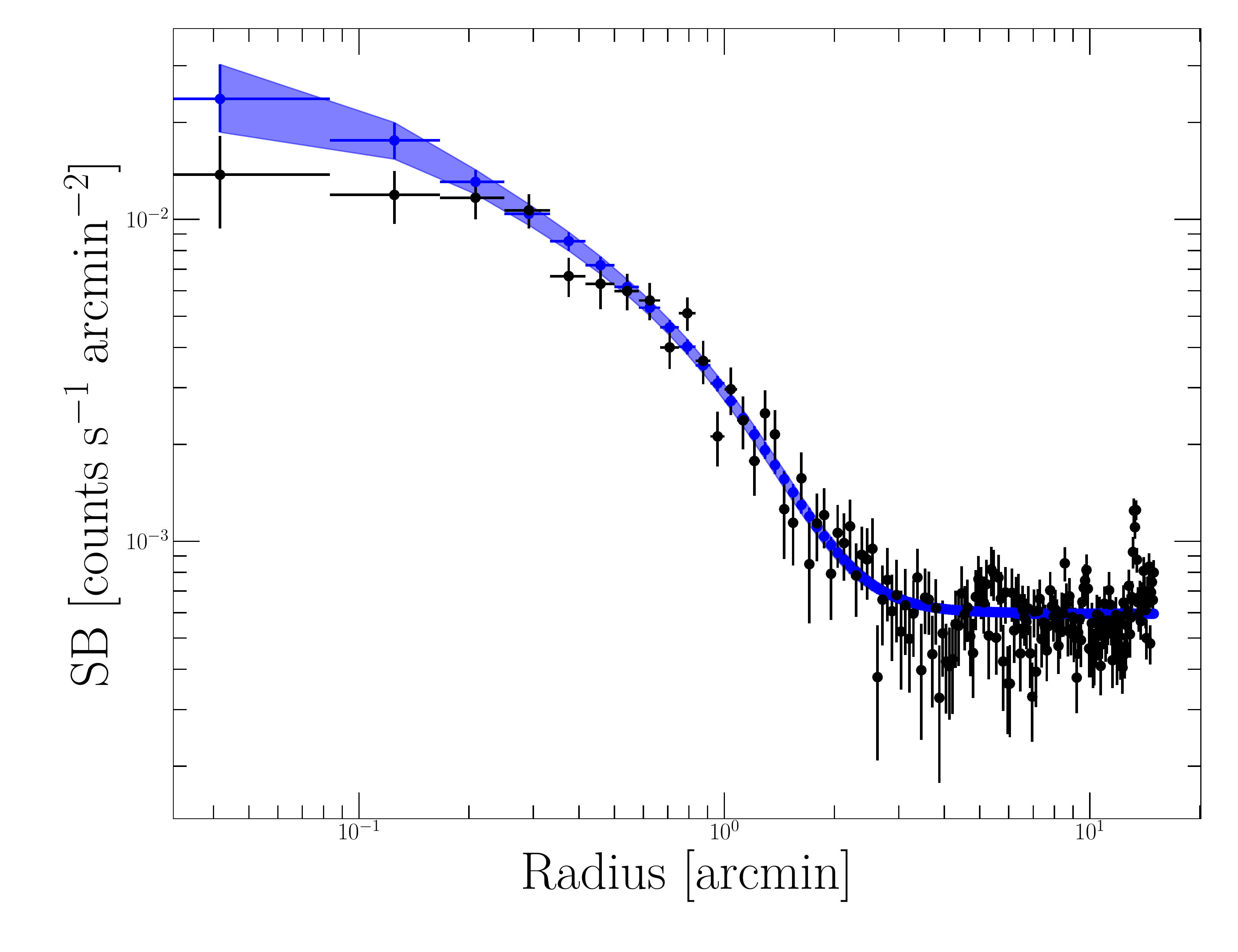}\label{profile_SB_n0234}} \hfill
   \subfloat[][]{\includegraphics[width=0.33\textwidth,height=4.52cm,trim= 0cm 0.15cm 0cm 0.35cm]{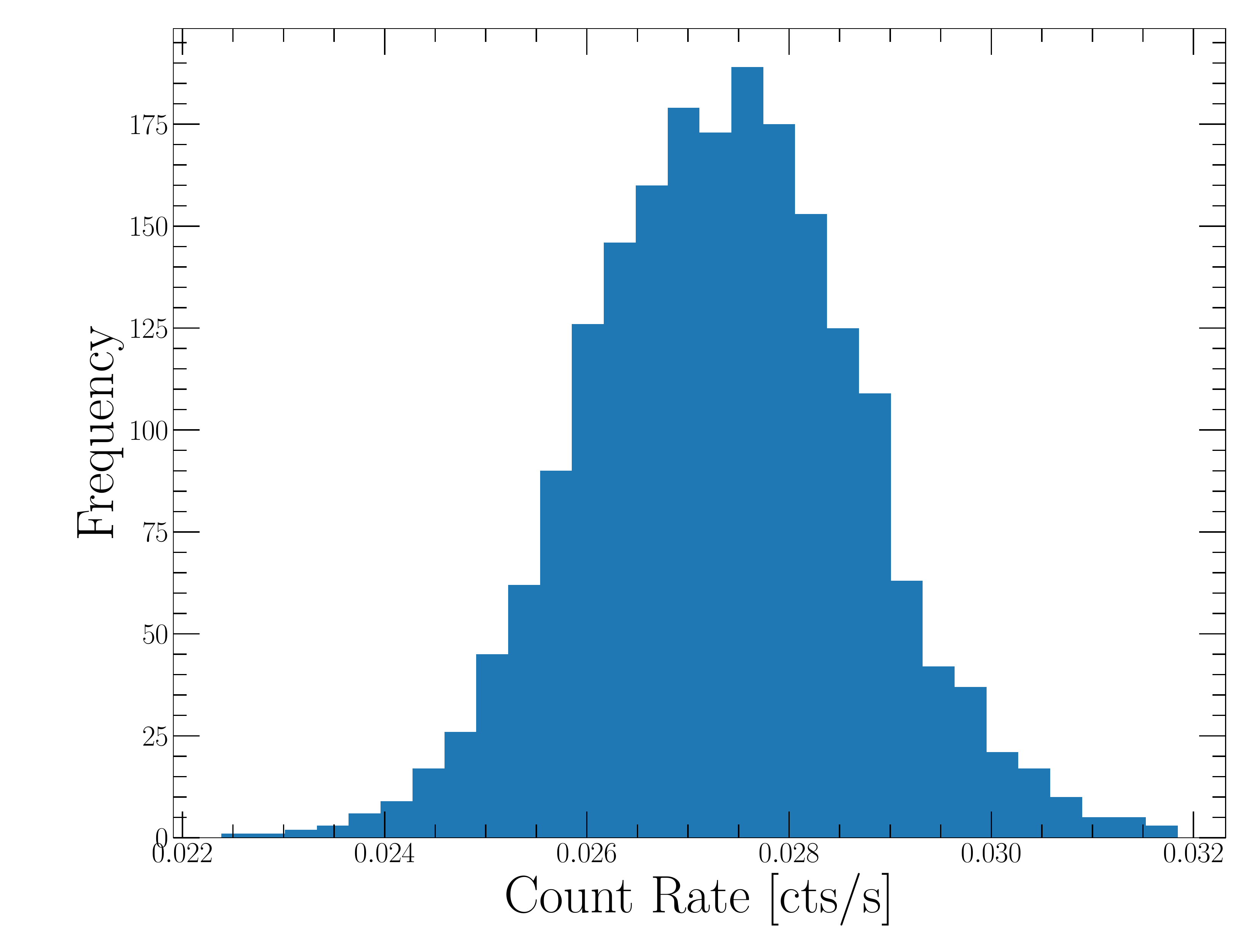}\label{countrate_dist_n0234}}\hfill
   \caption{Example of cluster count-rate measurement with the {\tt pyproffit} method. The displayed cluster is XLSSC 093 at a redshift of 0.429.
\newline
 (a) X-ray mosaic around the cluster; the image is 30 arcmin aside. (b) Particle background map. (c)  Combined exposure map along with the masks hiding the  neighbouring sources.  (d) Extracted cluster profile (black crosses); the green line displays the particle background level extracted from  map (b): this component is already subtracted from the displayed profile. (e) Overlaid on the extracted profile, the reconstructed (PSF-deconvolved) profile is shown in blue along with the 1-$\sigma$ estimated uncertainty. (f) Count-rate posterior distribution (MOS1 normalised) of the reconstructed profile.}
              \label{measurement}
\end{figure*}

\subsection{Step 2: The $\theta_{c}$ measurements}

Given that cluster angular sizes directly intervene in the selection (Fig. \ref{selfunc}), we need to determine the apparent core radii along with associated uncertainties. This is performed under the hypothesis of a $\beta =2/3$ model, as used to determine the selection function by means of simulations \citep{Pacaud2006}. The convolution by the PSF assumes a mean function over the entire mosaic (a cluster is generally seen at several off-axis positions on different observations):

\begin{equation}
    \mathrm{PSF}(\theta_{on-axis}) = \left[1+\left(\frac{\theta_{on-axis}}{4.765''}\right)^2\right]^{-1.505}
\end{equation}

with, $\theta_{on-axis}$ the angular radius (on-axis). The fit is performed for each cluster on the [0.5-2] keV co-added XMM mosaic and includes the corresponding exposure and background maps; the neighbouring sources are removed from the data using the revised masks. 
We use \textit{i}minuit \citep{Dembinski2020} to fit the Cash log-likelihood \citep{Cash1979}.

\subsection{Step 3: Accurate count-rate measurements}
\label{proffit}
Last step is to obtain best count-rate measurements along with realistic uncertainties, for the three X-ray bands involved in the construction of the XOD (Sec.~\ref{sec:cluster_sample}).
To optimise the determination, we relax the hypothesis of a single $\beta =2/3$ model and rather model the cluster emission by a linear combination of $\beta$-profiles, following the {\tt pyproffit} methodology \citep{Eckert2020}:
\begin{equation}
    \mathrm{SB}(\theta) = \sum_{i=1}^{N_{\mathcal{F}}} \alpha_i \mathcal{F}_i(\theta)
    \label{pyprofile}
\end{equation}
with $N_{\mathcal{F}}$ is the total number of functions and $\alpha_i$ the model coefficients. $\{\mathcal{F}_i\}$ is defined by \cite{Eckert2016},
\begin{equation}
    \mathcal{F}_i(\theta) = \left[1+\left(\frac{\theta}{\theta_{c,i}}\right)^2\right]^{-3\beta_i+1/2}
    \label{pyprofile_2}
\end{equation}
where $\theta$ is the distance to the cluster center. 
The mean value of predicted counts ($\lambda$) in profile annulus $a$ is given by
\begin{equation}
    \lambda_a = \mathrm{PSF} \ast \left( A_a t_a \sum_{i=1}^{N_{\mathcal{F}}} \alpha_i \mathcal{F}_i(\theta_a) \right) + B_a 
\end{equation}
Practically, we consider six $\beta$ values within [0.6-3] and $N = \theta_{bkg}/5''$ values for $\theta_c$.
The set of coefficients $\{\alpha_i\}$ in equation \ref{pyprofile} that maximises the log-likelihood given by equation \ref{loglike_pypro} 
\begin{equation} \label{loglike_pypro}
   - \log(\mathcal{L}) = \sum_{a=1}^{3N/2} \mu_a - N_{c,a}\ \log(\lambda_a) 
\end{equation}
 will describe the best fitted profile. \\
Eventually, we obtain the reconstructed profile  by drawing 1,000 posterior samples using No-U-Turn Sampler, NUTS \citep{Hoffman2011}. From this profile sample, we draw the count-rate posterior distribution. The various steps of the whole procedure is illustrated in Fig.~\ref{measurement}; for further technical details we refer the reader to \citep{Eckert2020}.

\section{Likelihood}\label{like_sec_appendix}
\subsection{Probability distribution of galaxy clusters in the $(z,{\rm CR},{\rm HR})$ space}

We describe the distribution of X-ray clusters as a Poisson realisation of an underlying
continuous field. Thus, the number of observed clusters $\hat N_i$ in the bin $i$,
in the 3D space $(z_i,{\rm CR}_i,{\rm HR}_i)$ (or any other set of observables) follows the Poisson probability distribution 
\be
P_{\hat N_i} = \frac{\hat n_i^{\hat N_i}}{\hat N_i !} e^{-\hat n_i} ,
\label{eq:Poisson-def}
\ee
where $\hat n_i$ is the continuous cluster density field. This continuous field $\hat n_i$ is also a random variable because of the sample variance, following the fluctuations of the dark matter density field averaged over the survey volume.
It is given by
\be
\hat n_i = \int_{z_{i-}}^{z_{i+}} dz \frac{d\chi}{dz} {\cal D}^2 \int d{\vec\Omega} \int
d\ln M \frac{d\hat n}{d\ln M} \Theta_i[M,z] 
\label{eq:hat-n-i-def}
\ee
with
\begin{equation}
\begin{aligned}
\Theta_i[M,z] = & \int_{-\infty}^{\infty} \frac{d\epsilon_{\rm CR}}{\sqrt{2\pi} \sigma_{\rm CR}}
e^{-\epsilon^2_{\rm CR}/(2 \sigma^2_{\rm CR})} 
\int_{-\infty}^{\infty} \frac{d\epsilon_{\rm HR}}{\sqrt{2\pi} \sigma_{\rm HR}} \\
& \times e^{-\epsilon^2_{\rm HR}/(2 \sigma^2_{\rm HR})} \\
& \times \Theta( {\rm CR}_{i-} <  f_{\rm CR}(M,z)+\epsilon_{\rm CR} < {\rm CR}_{i+} ) \\
& \times \Theta( {\rm HR}_{i-} <  f_{\rm HR}(M,z)+\epsilon_{\rm HR} < {\rm HR}_{i+} )
\label{eq:Theta-i-def}
\end{aligned}
\end{equation}
Here, $f_{\rm CR}(M,z)$ is the count rate associated with a cluster of mass $M$ at redshift $z$, with a Gaussian scatter $\sigma_{\rm CR}$. In practice, we can use a lognormal scatter by considering $\ln {\rm CR}$ as our observable, or by replacing the Gaussian integral in (\ref{eq:Theta-i-def}) by a lognormal distribution. Then, the first factor $\Theta$ is a unit top-hat that is nonzero when the count rate
falls in the bin $i$. Similar notations are used for the hardness ratio. Thus, $\Theta_i[M,z]$ is the probability that a cluster of mass $M$ at redshift $z$ falls
in the 2D bin $({\rm CR}_i,{\rm HR}_i)$. In Eq.(\ref{eq:hat-n-i-def}) we integrate the number of clusters over the redshift bin $\Delta z_i = z_{i+}-z_{i-}$, the survey angular area $\Delta \Omega$ and the cluster mass $M$, where $\chi$ and ${\cal D}$ are the radial and angular comoving distances and $\frac{d\hat n}{d\ln M}$ is the observed cluster mass function. Selection effects are included in the mass function $\frac{d\hat n}{d\ln M}$, which differs from the halo mass function and contains the response of the instrument.\\
\newline
At the level of the continuous field, the mean number of clusters in the bin $i$ is
\be
\bar n_i \equiv \langle \hat n_i \rangle 
= \Delta\Omega \int_{z_{i-}}^{z_{i+}} dz \frac{d\chi}{dz} {\cal D}^2 
\int d\ln M \frac{d n}{d\ln M} \Theta_i[M,z] ,
\label{eq:n-i-def}
\ee
where $\frac{d n}{d\ln M}$ is the cluster mass function predicted by a given cosmological scenario (including the selection effects).
If we neglect sample variance, we take $\hat n_i = \bar n_i$ without any scatter,
and the Poisson distribution (\ref{eq:Poisson-def}) has the fixed mean $\bar n_i$.
This provides the shot-noise contribution to the measurement error bars, associated
with the discreteness of the cluster distribution.\\
\newline
To estimate the impact of the sample variance, we consider the covariance of the continuous number counts $\hat n_i$. We have
\begin{equation}
\begin{aligned}
C_{ij} & \equiv \langle \hat n_i \hat n_j \rangle - \langle \hat n_i \rangle \langle \hat n_j \rangle\\
 & =  \int d\chi_1 d{\vec\Omega}_1 d\ln M_1 \int d\chi_2 d{\vec\Omega}_2 d\ln M_2  
{\cal D}_1^2 {\cal D}_2^2 \Theta_i[M_1,z_1] \\
& \times \Theta_j[M_2,z_2] \frac{dn}{d\ln M}(M_1,z_1)  \frac{dn}{d\ln M}(M_2,z_2) 
\xi_{12}({\vec x}_1-{\vec x}_2) \\
&
\end{aligned}
\end{equation}
where $\xi_{12}$ is the two-point correlation of halos of mass $M_1$ and $M_2$.
We assume that the redshift bins are much larger than the correlation length of the clusters and that the correlation function can be factorised as
\be
\xi_{12}(x) = b_1 b_2 \xi(x) ,
\label{eq:xi-ij-def}
\ee
where $\xi(x)$ is the dark matter correlation function and $b(M,z)$ is the bias of clusters of mass $M$ at redshift $z$. We use the \cite{Tinker2010} bias model during this analysis. Then, neglecting finite size effects associated with the borders of the survey volume, we can write the covariance matrix as
\be
C_{ij} = \delta_{z_i,z_j} \bar n_i \bar b_i \bar n_j \bar b_j \bar\xi_i ,
\label{eq:Cij-nb-nb-xi}
\ee
where $\delta_{z_i,z_j}$ is the Kronecker symbol with respect to the redshift bins $i$ and $j$, $\bar\xi_i$ the mean correlation in the redshift bin $i$, defined by
\be
\bar\xi_i = \int_{\chi_{i-}}^{\chi_{i+}} \frac{d\chi}{\Delta\chi_i} 
\int \frac{d{\vec\Omega}_1 d{\vec\Omega}_2}{(\Delta\Omega)^2} \xi({\vec x}_1-{\vec x}_2) ,
\label{eq:xi-def}
\ee
and $\bar b_i$ the mean bias defined by
\be
\bar n_i \bar b_i = \Delta\Omega \int_{z_{i-}}^{z_{i+}} dz \frac{d\chi}{dz} {\cal D}^2 
\int d\ln M \frac{d n}{d\ln M} \Theta_i[M,z] b(M,z) .
\label{eq:b-i-def}
\ee
\newline
Because of the Kronecker redshift factor in the covariance matrix (\ref{eq:Cij-nb-nb-xi}), different redshift bins are decoupled and we can analyse each redshift bin separately. Therefore, in the following we focus on a single redshift bin and the index $j$ refers only to the 2D bins $({\rm CR}_j,{\rm HR}_j)$.
It is also useful to consider the total number $\hat N$ of clusters in the survey and its continuous counterpart $\hat n$. For non-overlapping bins $j$ we have
\be
\hat N = \sum_j \hat N_j , \;\;\; \hat n = \sum_j \hat n_j , \;\;\;
\bar n = \sum_j \bar n_j , \;\;\; \bar n \bar b = \sum_j \bar n_j \bar b_j .
\label{eq:N-tot-def}
\ee
Let us define the fluctuations $\hat\delta_j$ and $\hat\delta$ of the continuous number counts $\hat n_j$ and $\hat n$ in the bin $j$ and in the full 2D volume $({\rm CR},{\rm HR})$,
\be
\hat n_j = (1+\hat\delta_j) \bar n_j , \;\;\; \hat n = (1+\hat\delta) \bar n , 
\ee
with means and covariances
\be
\langle \hat\delta_j \rangle = 0 , \;\;\; 
\langle \hat\delta_j \hat\delta_\ell \rangle =  \bar b_j \bar b_\ell \bar\xi , \;\;\;
\langle \hat\delta \rangle = 0 , \;\;\; 
\langle \hat\delta_j \hat\delta \rangle =  \bar b_j \bar b \bar\xi ,
\ee
and unit correlation coefficients
\be
\frac{\langle\hat\delta_j\hat\delta_\ell\rangle}{ \langle\hat\delta_j^2\rangle^{1/2}
\langle\hat\delta_\ell^2\rangle^{1/2}} = 1 , \;\;\;
\frac{\langle\hat\delta_j\hat\delta\rangle}{ \langle\hat\delta_j^2\rangle^{1/2}
\langle\hat\delta^2\rangle^{1/2}} = 1.
\ee
This implies that the fluctuations $\{\hat\delta_j,\hat\delta\}$ are linear functions of each other, and we obtain
\be
\hat\delta_j = \frac{\bar b_j}{\bar b} \hat\delta .
\label{eq:delta-i-bi-b-delta}
\ee
This is a consequence of the factorisation (\ref{eq:xi-ij-def}) of the cluster correlation function. We also note $\sigma_\delta^2$ the variance of the total number density contrast in the redshift bin,
\be
\sigma_\delta^2 \equiv \langle \hat\delta^2 \rangle = \bar b^2 \bar\xi .
\label{eq:sigma2-delta-def}
\ee

\subsection{Mean correlation $\bar\xi$}

For small angular windows and large enough redshift bins, it is possible to simplify the computation of the mean correlation $\bar \xi$ defined in Eq.(\ref{eq:xi-def}).
In the flat-sky approximation, for circular windows of angular radius $\theta_s$,
it reads as
\be
\bar\xi = \int \frac{d\chi}{\Delta\chi} \int \frac{d{\vec\theta}d{\vec\theta'}}{(\pi\theta_s^2)^2}
\int d{\vec k} \, e^{i k_\parallel (\chi-\chi_0)+i {\vec k}_{\perp}\cdot{\cal D} ({\vec\theta}'-{\vec\theta})}
P(k,z_0) ,
\ee
where $\chi_0$ is the comoving radial distance to the median redshift $z_0$
of the bin and $P(k,z_0)$ is the matter density power spectrum at redshift $z_0$.
For redshift bins that are not too shallow, $\Delta\chi \gg {\cal D}\theta_s$,
the integral over $\chi$ along the line of sight suppresses the contributions from
parallel wave numbers $k_{\parallel} > 1/(\Delta\chi)$, so that $\bar\xi$ is dominated by transverse wave numbers $k_\perp \sim 1/({\cal D}\theta_s) \gg k_{\parallel}$ and $k \simeq k_\perp$. This is Limber's approximation in its Fourier form. Then, the integral over $\chi$ gives a Dirac factor $2\pi\delta_D(k_\parallel)$, and the integration over $k_{\parallel}$ yields
\be
\bar\xi = \frac{2\pi}{\Delta\chi} \int \frac{d{\vec\theta}d{\vec\theta'}}{(\pi\theta_s^2)^2}
\int d{\vec k}_\perp \, e^{i {\vec k}_{\perp}\cdot{\cal D} ({\vec\theta}'-{\vec\theta})}
P(k_\perp) .
\ee
Introducing the 2D Fourier-space circular window $W_2(k_\perp {\cal D} \theta_s)$,
\be
W_2(k_\perp {\cal D} \theta_s) = \int \frac{d{\vec\theta}}{\pi\theta_s^2}
e^{i {\vec k}_{\perp}\cdot{\cal D} {\vec\theta}} 
= \frac{2 J_1(k_\perp {\cal D} \theta_s)}{k_\perp {\cal D} \theta_s} ,
\ee
where $J_1$ is the Bessel function of first order and first type, we obtain
\be
\bar\xi = \frac{4 \pi^2}{\Delta\chi} \int_0^{\infty} k\ dk\  P(k)
W_2(k_\perp {\cal D} \theta_s)^2 .
\ee

\subsection{Likelihood for the number counts in $({\rm CR},{\rm HR})$}

We now extend the likelihood-ratio analysis of Cash (1979) to our case.
Denoting $\theta_\alpha$ the parameters of the model, such as the set of 
cosmological parameters and additional cluster parameters, we consider the likelihood $L_{z_i}(\theta_\alpha;\hat N_j)$ in the redshift bin $z_i$ defined by
\be
L_{z_i}(\theta_\alpha;\hat N_j) =  P(\hat N_j;\theta_\alpha) = 
\int d\delta \, P(\delta) \prod_j P_{\hat N_j} .
\ee
Here we used the fact that the Poisson probabilities $P_{\hat N_j}$, defined in 
Eq.(\ref{eq:Poisson-def}), are governed by the continuous number counts $n_j$, which
are characterised by their means $\bar n_j$ and the fluctuating part $\delta$ from
Eq.(\ref{eq:delta-i-bi-b-delta}). The means $\bar n_j$, the bias $\bar b_j$ and the variance $\sigma_\delta^2$ themselves depend on the cosmological parameters
$\theta_\alpha$. At the level of the second-order moment for $\delta$, assuming the survey is large enough so that the relative fluctuations $\delta$ of the total number of clusters in a redshift bin are small, we take $P(\delta)$ to be Gaussian so that its probability distribution is fully determined by its variance. This yields
\ba
L_{z_i}(\theta_\alpha;\hat N_j) & = & \int_{-\infty}^{\infty} \frac{d\delta}{\sqrt{2\pi}\sigma_\delta} 
e^{-\delta^2/(2\sigma_\delta^2)} \prod_j 
\frac{[(1+\frac{\bar b_j}{\bar b} \delta) \bar n_j ]^{\hat N_j}}{\hat N_j !} \nonumber \\
&& \times e^{- (1+\frac{\bar b_j}{\bar b} \delta) \bar n_j } ,
\ea
which also reads as
\ba
L_{z_i}(\theta_\alpha;\hat N_j) & = & \prod_j \frac{\bar n_j^{\hat N_j}}{\hat N_j !} 
e^{-\bar n_j} \int_{-\infty}^{\infty} \frac{d\delta}{\sqrt{2\pi}\sigma_\delta} 
e^{-\delta^2/(2\sigma_\delta^2)} \nonumber \\
&& \times \; e^{\sum_j \hat N_j \ln(1+\frac{\bar b_j}{\bar b} \delta) - \bar n \delta } .
\label{eq:Lzi-delta}
\ea
Here we used the last property in (\ref {eq:N-tot-def}). The first product, independent of $\delta$, is the usual shot-noise contribution, whereas the
integral over $\delta$ of the second product gives the contribution from the sample variance. If the latter is negligible, $\sigma_{\delta}\to 0$, it goes to unity and we recover the shot-noise value. If the volume is large enough, the relative fluctuation $\sigma_\delta$ of the total number of clusters is small and we can expand the logarithm up to second order in $\delta$,
\be
\ln\left(1+\frac{\bar b_j}{\bar b} \delta\right) = \frac{\bar b_j}{\bar b} \delta 
- \frac{1}{2} \frac{\bar b_j^2}{\bar b^2} \delta^2 + \dots
\label{eq:ln-1+delta}
\ee
This is valid if we have
\be
\sum_j \hat N_j \frac{1}{3} \frac{\bar b_j^3}{\bar b^3} \delta^3 \sim \frac{(\delta N)^3}{3 N^2} \ll 1,
\ee
where we wrote $\delta = \delta N/N$ and $\bar b_j \sim \bar b$. For large survey sizes, with $N \gg 1$, we typically expect $\delta N \sim \sqrt{N}$ so that the
approximation (\ref{eq:ln-1+delta}) is valid. Then, we can perform the Gaussian integration in Eq.(\ref{eq:Lzi-delta}), which gives
\begin{equation}
\begin{aligned}
& L_{z_i}(\theta_\alpha;\hat N_j) = \prod_j \frac{\bar n_j^{\hat N_j}}{\hat N_j !} 
e^{-\bar n_j} \left( 1+\sigma^2_\delta \sum_j \hat N_j \frac{\bar b_j^2}{\bar b^2} \right)^{-1/2}\\
& \times \, \exp\left[ \frac{\sigma_\delta^2}{2} \left( \sum_j \hat N_j \frac{\bar b_j}{\bar b} - \bar n \right)^2
\left( 1+\sigma^2_\delta \sum_j \hat N_j \frac{\bar b_j^2}{\bar b^2} \right)^{-1} \right] . 
\label{eq:Lzi-1}
\end{aligned}
\end{equation}
For the estimation of the cosmological and cluster parameters $\theta_\alpha$ with the likelihood method (Cash 1979), we compare the logarithm ${\cal L}=-\ln L$ obtained for different sets of parameters. The estimated parameters $\theta_\alpha^{\rm obs}$ are those that minimise ${\cal L}$ and the variation of ${\cal L}$ with $\theta_\alpha$ provides the confidence intervals, following a $\chi^2$ law. Thus, we consider
\begin{equation}
\begin{aligned}
{\cal L}_{z_i}(\theta_\alpha;\hat N_j) = & \bar n - \sum_j \hat N_j \ln(\bar n_j) 
\\
& + \frac{1}{2} \ln\left( 1+\sigma^2_\delta \sum_j \hat N_j \frac{\bar b_j^2}{\bar b^2} \right)\\
& - \frac{\sigma_\delta^2}{2} \left( \sum_j \hat N_j \frac{\bar b_j}{\bar b} - \bar n \right)^2 \times \left( 1+\sigma^2_\delta \sum_j \hat N_j \frac{\bar b_j^2}{\bar b^2} \right)^{-1} ,
\label{eq:lnLzi-1}
\end{aligned}
\end{equation}
where we used $\sum_j \bar n_j = \bar n$, within a given redshift bin. We discarded as usual the term $\ln(\hat N_j !)$, because it does not depend on the parameters $\theta_\alpha$ and cancels out in the difference ${\cal L}(\theta_\alpha) - {\cal L}(\theta_\alpha')$. The first term is the usual shot-noise contribution while the other two terms are the sample-variance contribution, which vanishes for $\sigma_\delta\to 0$. The sums over $j$ only need to run over the bins in the 2D space $({\rm CR},{\rm HR})$ that are not empty, as they come with a factor $\hat N_j$. This ensures that the results are not affected if we enlarge the box in the 2D space $({\rm CR},{\rm HR})$ to a large volume far beyond the realistic domain, including regions that are always empty.\\
\newline
Going back to the 3D space $(z,{\rm CR},{\rm HR})$, because the redshift bins are independent we simply have for the full likelihood
\be
L = \prod_i L_{z_i} , \;\;\; {\cal L} = \sum_i {\cal L}_{z_i} .
\ee
\newline
In practice, the parameters $\theta_\alpha$ should not be far from those derived from previous experiments, such as Planck (for the cosmological parameters). Then, as in Fisher matrix analysis where we neglect the cosmological dependence of the covariance matrix, we can neglect the dependence on $\theta_\alpha$ of the sample-variance quantities $\{\sigma_\delta^2,\bar b_j, \bar b\}$, which we compute for a reference cosmology labeled by the subscript $(0)$, $\{\sigma^2_{\delta(0)},\bar b_{j(0)}, \bar b_{(0)} \}$. Then, we test the cosmological scenario through its predictions for the means $\bar n_j$. This implies that we can discard the second factor in Eq.(\ref{eq:lnLzi-1}), as it does not depend on $\theta_\alpha$, and write
\begin{equation}
\begin{aligned}
{\cal L}_{z_i}(\theta_\alpha;\hat N_j) = & \bar n - \sum_j \hat N_j \ln(\bar n_j) 
- \frac{\sigma^2_{\delta(0)}}{2} \\
& \times \left( \sum_j \hat N_j \frac{\bar b_{j(0)}}{\bar b_{(0)}} 
- \bar n \right)^2 \left( 1+\sigma^2_{\delta(0)} \sum_j \hat N_j 
\frac{\bar b^2_{j(0)}}{\bar b^2_{(0)}} \right)^{-1}  \hspace{0.5cm} 
\label{eq:lnLzi-2}
\end{aligned}
\end{equation}

\subsection{Behaviour of the likelihood ${\cal L}$}
\subsubsection{Cosmology selected by the data}

\begin{figure}
   \centering
   \includegraphics[width=0.45\textwidth]{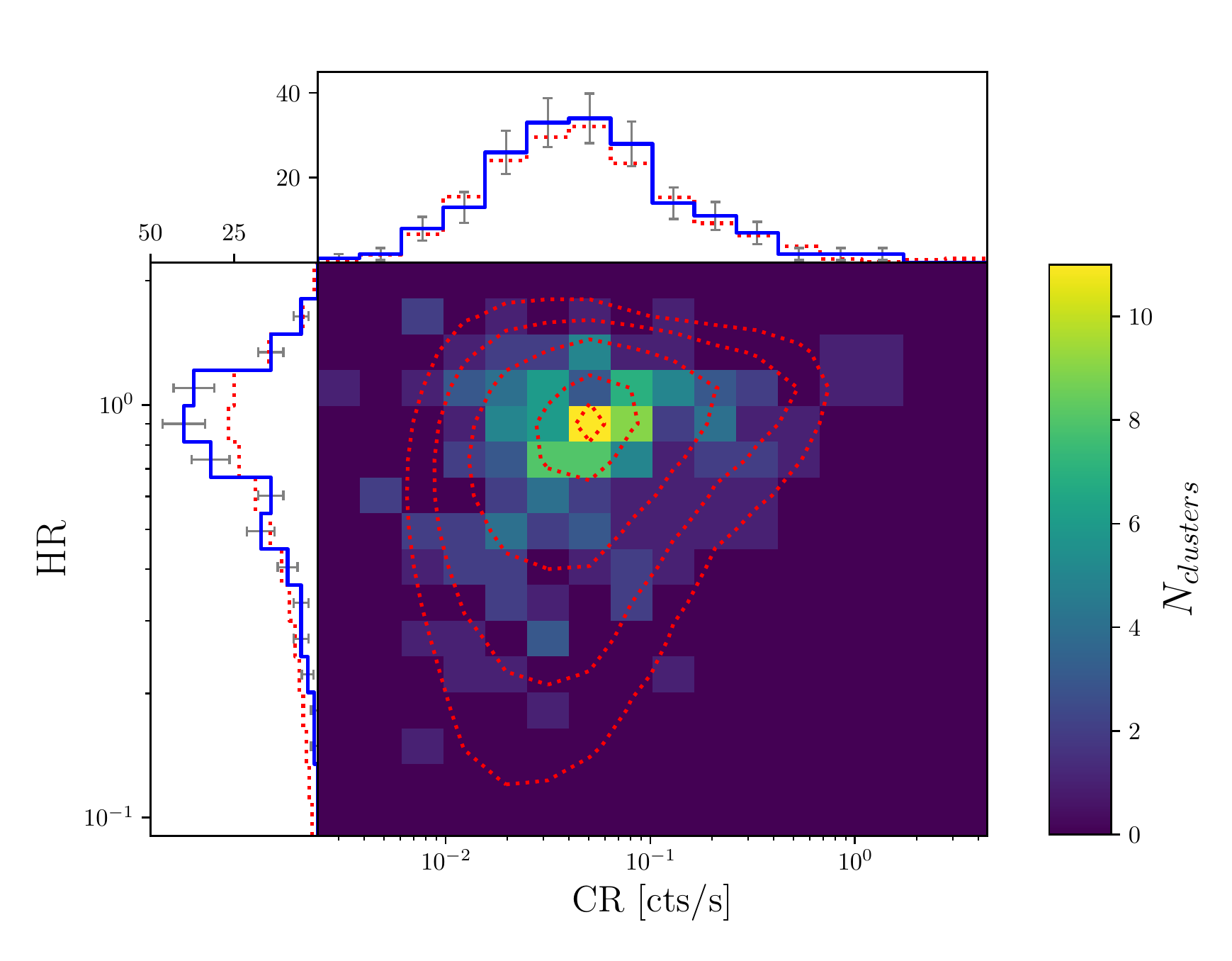} \hfill
   \caption{The X-ray observable diagram (XOD) of the XXL C1 sample, integrated over the redshift range [0.05-1], used in this study, together with the 1D CR, HR distributions in blue. In red, the theoretical diagram predicted by our $\mathrm{\Lambda}$CDM best fit parameters given in section \ref{main_results}. We can see where the CR-HR space is not dominated by the shot noise ($\hat N_j$ > 2-4 objects), we obtain, in average, that $\bar n_j$ is smaller than $\hat N_j$. Error bars only account for shot noise.}
   \label{Diag_178_with_contours}%
\end{figure}

Let us now investigate the behaviour of ${\cal L}_{z_i}$ as a function of the theoretical means $\bar n_j$. This will provide us some insight into the response of the cosmological parameters $\theta_\alpha$ to the measurements $\hat N_j$, through the associated means $\bar n_j$. Thus, the cosmology selected by the measurement corresponds to the set $\{ \bar n_j \}$ that maximises the likelihood $L_{z_i}$, i.e. that minimises the negative logarithm ${\cal L}_{z_i}$. For ${\cal N}$ 2D bins at redshift $z_i$, this gives the ${\cal N}$ equations $\frac{\partial{\cal L}}{\partial \bar n_j}=0$,
\be
1 \leq j \leq {\cal N} : \;\;   1 - \frac{\hat N_j}{\bar n_j} + \sigma^2 \sum_\ell 
\left( \hat N_\ell \frac{\bar b_{\ell(0)}}{\bar b_{(0)}} - \bar n_\ell \right) = 0 ,
\label{eq:minimum-L}
\ee
where we used $\bar n = \sum_j \bar n_j$ and we defined
\be
\sigma^2 = \sigma^2_{\delta(0)} \left( 1+\sigma^2_{\delta(0)} \sum_j \hat N_j 
\frac{\bar b^2_{j(0)}}{\bar b^2_{(0)}} \right)^{-1} > 0 .
\label{sigma-def}
\ee 
First, we note that if the measurements are equal to the reference predictions
we recover the reference cosmology:
\be
\mbox{if } \hat N_j = \bar n_{j(0)} \mbox{ then } \bar n_j = \bar n_{j(0)} ,
\label{eq:Nj-nj0}
\ee
where we used the sum rules (\ref{eq:N-tot-def}).\\
Second, we note that the ${\cal N}$ equations (\ref{eq:minimum-L}) admit the solution
\be
\bar n_j = \alpha \hat N_j ,
\ee
where $\alpha$ is a solution of the single equation
\be
1 - \frac{1}{\alpha} + \sigma^2 ( \hat N_{(b)} - \alpha \hat N ) = 0 ,
\label{eq:alpha-def}
\ee
where we introduced
\be
\hat N_{(b)} =  \sum_j \hat N_j \frac{\bar b_{j(0)}}{\bar b_{(0)}} .
\ee
We can understand this from the fact that sample-variance fluctuations of the total 
number of clusters do not affect the relative counts in the different pixels 
$({\rm CR}_i,{\rm HR}_i)$, as seen in Eq.(\ref{eq:delta-i-bi-b-delta}). Therefore, the inferred ratios $\bar n_j/\bar n_\ell$ are equal to the measured ratios $\hat N_j/\hat N_\ell$. Thus, this likelihood method selects cosmologies that predict the observed distribution profile in the 2D space $({\rm CR},{\rm HR})$, up to a uniform rescaling $\alpha$.\\
Next, the quadratic equation (\ref{eq:alpha-def}) has two solutions,
\be
\alpha_\pm = \frac{1+\sigma^2 \hat N_{(b)} \pm 
\sqrt{ ( 1+\sigma^2 \hat N_{(b)} )^2 - 4 \sigma^2 \hat N }} {2 \sigma^2 \hat N} ,
\ee
with the asymptotic behaviours for $\sigma^2 \to 0$,
\be
\alpha_- \simeq 1 + \sigma^2 (\hat N - \hat N_{(b)}) + \dots , \;\;
\alpha_+ \simeq \frac{1}{\sigma^2 \hat N} \to \infty .
\label{eq:alpha-low-sigma2}
\ee
The physical solution is $\alpha_-$, which goes to unity when the sample variance is negligible and we recover the shot-noise likelihood, where the inferred cosmological values are $\bar n_j = \hat N_j$.\\
The second solution $\alpha_+$ is not physical and is due to the approximations in our treatment, such as (\ref{eq:ln-1+delta}). Indeed, it would correspond to a large uniform density fluctuation $\delta \simeq -1$, where the approximation (\ref{eq:ln-1+delta}) is no longer valid. In practice $\alpha_+$ does not appear and does not impair the likelihood algorithm because we restrict the search in the cosmological parameter space to a small realistic region, not too far from the Planck values (i.e., we do not consider cosmologies that would predict
ten times more clusters or further than the Planck concordance cosmology).\\
\newline
We can see from Eq.(\ref{eq:alpha-def}) that $\alpha=1$ is a solution if $\hat N_{(b)} = \hat N$. This leads to a generalisation of the solution (\ref{eq:Nj-nj0}),
\be
\mbox{if } \hat N_j = \beta \bar n_{j(0)} \mbox{ then } \bar n_j = \hat N_j ,
\label{eq:nj-Nj-beta}
\ee
which applies for any $\beta > 0$. Therefore, if the measurements $\hat N_j$ follow the same 2D profile as the reference cosmology, up to a uniform multiplicative factor $\beta$, there is no rescaling and the likelihood method selects cosmologies that predict the same number counts $\bar n_j$ as those that are measured.\\
\newline
If $\hat N_{(b)} > \hat N$, we can see from Eq.(\ref{eq:alpha-low-sigma2}) that $\alpha < 1$. Therefore, the likelihood selects cosmologies that predict mean counts $\bar n_j$ that are lower than the measured values $\hat N_j$, and the required increase up to $\hat N_j$ is explained by a local positive fluctuation $\delta > 0$ of the density field, arising from a sample-variance effect. This can be understood from the fact that $\hat N_{(b)} > \hat N$ means that higher-bias pixels have a greater count than expected. This points towards a positive density fluctuations $\delta > 0$. This is similar to the well-known Kaiser derivation of the bias of rare objects like clusters, with respect to the underlying dark matter density field. There, using for instance the Press-Schechter mass function or the peak formalism, it is noticed that positive large-scale matter density fluctuations $\delta > 0$ enhance the formation of rare massive objects, and the more extreme the object (i.e. a larger mass) the greater the enhancement. This means a larger bias for more massive halos, due to the increased sensitivity of the large-mass tail of the mass function. Reversing this picture, we can see that enhanced number counts of rare massive halos, i.e. of high-bias objects, arise from positive fluctuations of the underlying matter density field.\\
Therefore, in our case (and more generally), $\hat N_{(b)} > \hat N$ signals
an enhancement of high-bias objects and hence a positive underlying density 
fluctuation $\delta>0$. To accommodate this amplification with the observed values $\hat N_j$, the means $\bar n_j$ must then be somewhat smaller than the targets $\hat N_j$.\\
In Fig.~\ref{Diag_178_with_contours}, we show the XXL C1 sample diagram together with the diagram predicted by our $\mathrm{\Lambda}$CDM best fit parameters given in section \ref{main_results}. While the diagram predicted by the best fit parameters follow the peak position and the shape of the observed diagram, we can see that, as expected, in the CR-HR space not dominated by the shot noise, $\bar n_j$ is smaller than $\hat N_j$.

\subsubsection{Confidence intervals}

\begin{figure}
   \centering
   \includegraphics[width=0.45\textwidth]{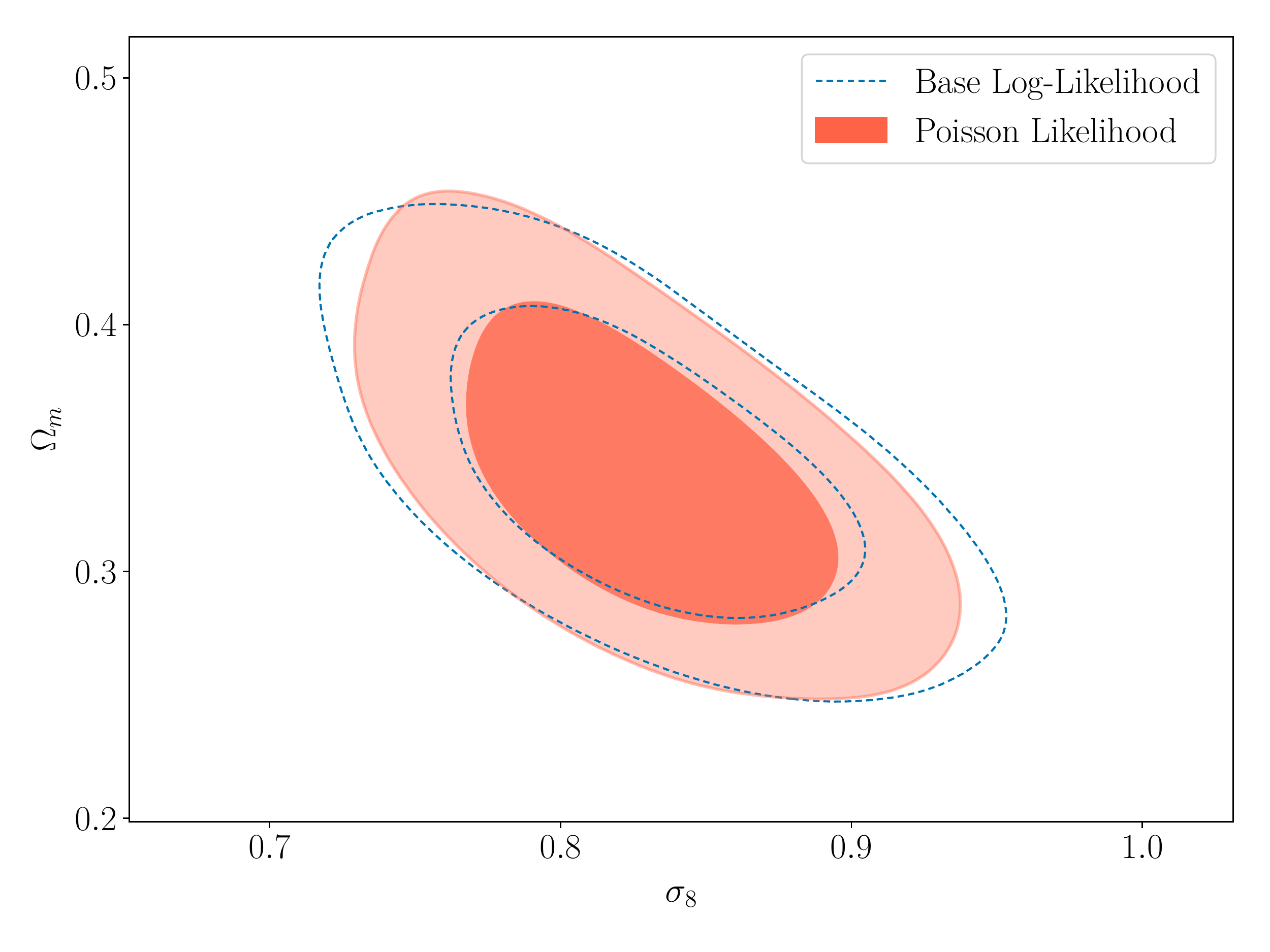} \hfill
   \caption{$\Omega_{m}-\sigma_{8}$ contours using the log-likelihood from equation \ref{eq:lnLzi-2} (Base Log-Likelihood) and the Poisson log-likelihood from equation \ref{eq:poissonlike} (Poisson Likelihood). The constraints when using the base log-likelihood compared to the Poisson one are increased by a factor of 10\%.}
              \label{like_stud}%
\end{figure}

We also expect the sample variance to increase the error bars obtained for the parameters $\theta_\alpha$ from the observations, as compared with the shot-noise-only estimate.\\
From Eq.(\ref{eq:lnLzi-2}) we obtain the Hessian
\be
H_{j\ell} \equiv \frac{\partial^2 {\cal L}}{\partial\bar n_j \partial \bar n_\ell} 
= \frac{\hat N_j}{\bar n_j^2} \delta_{j\ell} - \sigma^2 .
\ee
We clearly see that the sample-variance contribution decreases the curvature of the
likelihood ${\cal L}$ and therefore increases the size of the confidence interval.
For instance, we obtain for the trace and determinant of the Hessian
\be
{\rm Tr}(H) = \sum_j  \frac{\hat N_j}{\bar n_j^2} - {\cal N} \sigma^2 , 
\label{eq:Tr-H}
\ee
\be
{\rm det}(H) = \left( \prod_j  \frac{\hat N_j}{\bar n_j^2} \right) 
\left( 1 - \sigma^2 \sum_j \frac{\bar n_j^2}{\hat N_j} \right) ,
\label{eq:det-H}
\ee
which are decreased by the sample-variance term proportional to $\sigma^2$.
From Eqs.(\ref{eq:Tr-H})-(\ref{eq:det-H}), we can estimate the change $\delta\lambda_j$ of the eigenvalues $\lambda_j$ of the Hessian $H_{ij}$ due to the sample-variance term. With $\hat N_j \sim \bar n_j$, $\lambda_j \sim \lambda + \delta\lambda$, we write
\be
{\rm Tr}(H+\delta H) \sim {\cal N} \lambda \left( 1 + \frac{\delta\lambda}{\lambda} \right) ,
\ee
\be
{\rm det}(H+\delta H) \sim \lambda^{{\cal N}} \left( 1 + \frac{\delta\lambda}{\lambda} \right)^{\cal N}
\sim \lambda^{{\cal N}} \left( 1 + {\cal N} \frac{\delta\lambda}{\lambda} \right) .
\ee
For both the trace and the determinant, the comparison with Eqs.(\ref{eq:Tr-H})-(\ref{eq:det-H}) gives the order-of-magnitude estimate
\be
\frac{\delta\lambda}{\lambda} \sim - \sigma^2 \bar n_j ,
\label{eq:delta-lambda}
\ee
where $\bar n_j$ is the typical number count in a 2D cell. Then, we expect the interval of confidence on the cosmological parameters to increase by a factor of the order of $1+\sigma^2 \bar n_j/2$, when we include the effect of the sample variance. Here $\bar n_j$ should correspond to a binning that is well adapted to the survey, that is, which corresponds to the amount of information that can be drawn from
the observations. By choosing increasingly small bins one decreases $\bar n_j$
and the apparent magnitude of $\delta\lambda$ in Eq.(\ref{eq:delta-lambda}), but this is compensated by the larger size of the matrix $H_{j\ell}$ and the greater number of constraints, which are mostly degenerate.\\
\newline
For the likelihood (\ref{eq:lnLzi-2}) to be meaningful, the Hessian $H$ should be positive definite in the neighbourhood of the reference point $(0)$, so that the solutions of Eq.(\ref{eq:minimum-L}) correspond indeed to minima of ${\cal L}$, and not to local maxima or a saddle points. This requires the determinant (\ref{eq:det-H}) to be strictly positive. Substituting the expression (\ref{sigma-def}), we find that $\det(H)>0$ if
\be
1 + \sigma_{\delta(0)}^2 \sum_j \left( \hat N_j \frac{\bar b^2_{j(0)}}{\bar b^2_{(0)}} 
- \frac{\bar n_j^2}{\hat N_j} \right) > 0 .
\ee
This is positive when $\sigma_{\delta(0)}$ is small. Moreover, at the reference point, with $\hat N_j = \bar n_j = \bar n_{j(0)}$, this reads as
\be
1 + \frac{ \sigma_{\delta(0)}^2 }{ \bar b^2_{(0)} } \bar n \left( \langle \bar b^2_{j(0)} \rangle 
- \langle \bar b_{j(0)} \rangle^2 \right) > 0 ,
\ee
where we used the sum rules (\ref{eq:N-tot-def}) and we defined the averages
\be
\langle A_j \rangle = \sum_j p_j A_j \;\; \mbox{with} \;\; p_j = \frac{\bar n_{j(0)}}{\bar n_{(0)}} .
\ee
The weights $p_j$ are positive and sum to unity. Therefore, they can be interpreted 
as a probability distribution and we obtain $\langle \bar b^2_{j(0)} \rangle - \langle \bar b_{j(0)} \rangle^2 \geq 0$.\\
Thus, the determinant (\ref{eq:det-H}) is actually strictly positive at the reference point $\{ \bar n_{j(0)} \}$ for any value of $\sigma_{\delta(0)}$. By continuity on $\sigma_{\delta(0)}$, this also implies that the Hessian matrix $H$
is always positive definite at this reference point. If the search for the cosmological parameters does not go too far from this reference, the Hessian matrix always remains positive definite. This ensures that the likelihood (\ref{eq:lnLzi-2}) is well behaved.\\
The $\Omega_{m}-\sigma_{8}$ constraints from the Log-likelihood of Eq. (\ref{eq:lnLzi-2}) and a simple Poisson Log-Likelihood without sample variance contribution,
\begin{equation}\label{eq:poissonlike}
    {\cal L}_{z_i}(\theta_\alpha;\hat N_j) = \bar n - \sum_j \hat N_j \ln(\bar n_j) ,
\end{equation}
are shown in Fig. \ref{like_stud}.

\end{appendix}
\end{document}